\documentclass[fleqn,usenatbib]{mnras}

\usepackage{newtxtext,newtxmath}
\usepackage{float}
\usepackage{siunitx}
\usepackage{threeparttable}
\usepackage{subfig}
\usepackage{graphicx} 
\usepackage{xcolor}

\usepackage[T1]{fontenc}

\DeclareRobustCommand{\VAN}[3]{#2}
\let\VANthebibliography\thebibliography
\def\thebibliography{\DeclareRobustCommand{\VAN}[3]{##3}\VANthebibliography}

\usepackage{graphicx}	
\usepackage{amsmath}	

\title[Aliphatics' Effect on Aromatic Emission
]{The Influence of Aliphatic Components on the Aromatic Emission Characteristics of Polycyclic Aromatic Hydrocarbons
}

\author[Zhang \& Zhang]{
Zhuang Zhang,$^{1}$\
Yong Zhang$^{1,2}$\thanks{E-mail: zhangyong5@mail.sysu.edu.cn}
\\
$^{1}$School of Physics and Astronomy, Sun Yat-sen University, 2 Daxue Road, Zhuhai 519082, Guangdong Province, China\\
$^{2}$CSST Science Center for the Guangdong-Hong Kong-Macau Greater Bay Area, Sun Yat-Sen University, 2 Daxue Road, Zhuhai 519082,
Guangdong Province, China\\
}

\date{Accepted XXX. Received YYY, in original form ZZZ}

\pubyear{\the\year{}}

\begin{document}
\label{firstpage}
\pagerange{\pageref{firstpage}--\pageref{lastpage}}
\maketitle

\begin{abstract}
Intensity ratios of aromatic emission features are widely used to diagnose the size and ionization state of polycyclic aromatic hydrocarbons (PAHs) in astronomical environments. However, PAHs are known to typically carry aliphatic side chains, a structural feature that may compromise the reliability of traditional diagnostic methods. This study systematically investigates the effects of aliphatic components on the aromatic emission properties of PAHs. Based on theoretical data from the NASA Ames PAH IR Spectroscopic Database, we compare the emission behavior of purely aromatic PAHs with those containing aliphatic substituents, revealing that aliphatic functionalization 
may modify the intensity ratio of the 11.2 $\mu$m band relative to the 7.7 $\mu$m and 3.3 $\mu$m bands. 
This potentially leads to misidentification of their ionization state if molecular structural effects are neglected. Further analysis indicates that the impact of aliphatic components on diagnostic band ratios strongly depends on PAH size: small PAHs exhibit significant emission ratio shifts, deviating from traditional size/ionization trends, while larger PAHs are minimally affected. 
Despite these shifts, the classic \(I_{11.2/7.7}\) versus \(I_{11.2/3.3}\) diagnostic grid remains largely applicable to mixed aromatic-aliphatic PAHs, although some systematic calibration may be needed. Our findings emphasize the necessity for caution when interpreting PAH band ratios in aliphatic-rich environments, as variations in PAH molecular composition may distort inferences about physical conditions.
\end{abstract}




\section{Introduction}
Polycyclic aromatic hydrocarbons (PAHs) are widely recognized as primary carriers of the unidentified infrared emission features observed in the 3--20 $\mu$m wavelength range \citep[e.g.,][and references therein]{Tielens2008}. Since the pioneering studies by \citet{1984A&A...137L...5L} and \citet{1985ApJ...290L..25A}, which first attributed aromatic infrared (IR) bands to PAH molecules, PAHs have become a cornerstone for interpreting mid-IR spectra across diverse astrophysical environments. From a theoretical standpoint, the formation pathways of cosmic PAHs remain debated, with proposed channels including bottom-up synthesis in circumstellar envelopes of evolved stars and top-down processing/fragmentation of larger carbonaceous grains in the interstellar medium \citep[e.g.,][]{1989ApJ...341..372F,2012AA...542A..69P,2014NatCo...5.3054M}. The relative importance of these pathways may depend on local physical conditions and environmental metallicity \citep{2020NatAs...4..339L}. Recent detections of PAH derivatives in cold environments (e.g., 1-cyanopyrene in TMC-1) further suggest that relatively large PAHs (or their precursors) can assemble and survive at low temperatures, providing additional constraints on PAH formation and inheritance \citep{2024Sci...386..810W}.

Upon absorbing ultraviolet (UV) photons, PAH molecules undergo radiative cooling through molecular vibrational modes, producing a set of characteristic IR emission bands. Prominent features occur at 3.3, 6.2, 7.7, 8.6, and 11.2 $\mu$m, with physical origins commonly assigned to aromatic C--H stretching (3.3 $\mu$m), aromatic C--C stretching (6.2 $\mu$m), coupled aromatic C--C stretching and C--H in-plane bending (7.7 $\mu$m), aromatic C--H in-plane bending (8.6 $\mu$m), and aromatic C--H out-of-plane bending (11.2 $\mu$m) modes \citep{1977Natur.269..132K,1984A&A...137L...5L,langhoff1996theoretical,1981MNRAS.196..269D}. Recent high-resolution James Webb Space Telescope (JWST)  spectra further reveal that several of these bands comprise multiple blended components whose profiles and peak positions depend on PAH size, charge state, and molecular structure \citep{2024A&A...685A..75C,2025A&A...699A.133K}. In the 3--4 $\mu$m region, in addition to the aromatic 3.3 $\mu$m band, a weaker 3.4 $\mu$m complex is frequently observed and is commonly linked to aliphatic C--H stretches in methyl/methylene groups and superhydrogenated edges, while anharmonicity in the aromatic C--H stretching manifold may also contribute to the 3.3--3.4 $\mu$m substructure and red wing 
\citep{1996ApJ...472L.127B,2013ApJS..208...26S,2020ApJ...892...11B,2020ApJS..247....1Y,Mackie2022}.
Complementary aliphatic bending features at 6.85 and 7.25 $\mu$m have also been reported in some sources \citep{2016MNRAS.462.1551Y}, indicating that interstellar organic carriers may contain both aromatic and aliphatic components \citep{2011Natur.479...80K}.

The size and ionization state of PAHs are encoded in the relative strengths of the main PAH bands, motivating the widespread use of band-intensity ratios as diagnostic tools \citep[e.g.,][]{1989ApJS...71..733A,1993ApJ...415..397S,2001ApJ...551..807D}. In particular, the $I_{11.2}/I_{3.3}$ intensity ratio increases with PAH size and is widely used as a tracer for the characteristic size of predominantly neutral PAH populations \citep[e.g.,][]{2012ApJ...744...68M,2012ApJ...754...75R,2016A&A...590A..26C}. Ratios involving the 6--9 $\mu$m complex and the 11--12 $\mu$m bands are also sensitive to ionization and molecular structure, enabling two-dimensional diagnostic diagrams to infer size and charge distributions from observed spectra \citep[e.g.,][]{2001ApJ...551..807D}.

Recently, \citet{MPR2020} proposed a two-dimensional diagnostic diagram for PAH size and ionization state, using the intensity ratios $I_{11.2 + 11.0}/I_{7.7}$ and $I_{11.2 + 11.0}/I_{3.3}$ as tracers of ionization and molecular size, respectively. Based on a large set of model spectra from the NASA Ames PAH IR Spectroscopic Database\footnote{https://www.astrochemistry.org/pahdb} \citep{2010ApJS..189..341B,nasa2.00,nasa3.00ApJS23432B}, they mapped PAHs of different sizes and charge states within this diagnostic space and applied the method to galaxies and nebulae.
Subsequently, \citet{MPR2023} examined spectral variations under two extreme evolutionary scenarios (preferential removal of small versus large PAHs). They found that the $I_{11.2 + 11.0}/I_{7.7}$ ratio remains relatively stable, while the $I_{11.2 + 11.0}/I_{3.3}$ ratio varies monotonically with the average number of carbons and separates the two evolutionary pathways, reinforcing the utility of the $I_{11.2}/I_{3.3}$ ratio as a size tracer and as a potential indicator of dominant processing routes.

Despite these advances, most current diagnostic grids are calibrated predominantly using purely aromatic PAHs, with little systematic evaluation of how aliphatic moieties may bias inferred PAH properties. Observationally, the co-detection of the 3.4 $\mu$m feature alongside the aromatic 3.3 $\mu$m band suggests that PAH-related carriers are not exclusively aromatic and may contain a non-negligible fraction of aliphatic bonds \citep[e.g.,][]{1986ApJ...306L.105D,2007ApJ...662..389G,2014ApJ...784...53M,2022Ap&SS.367...16K}. In some environments the 3.4 $\mu$m emission can be comparable to the 3.3 $\mu$m band, motivating a quantitative evaluation of how such aliphatic contributions perturb commonly used size--charge diagnostics.

Laboratory and theoretical studies consistently show that aliphatic substituents can strongly modify the 3~$\mu$m C--H stretching region of PAHs \citep{2012ApJ...755..120M, MAURYA20151, 2018A&A...610A..65M}.
Quantum chemical calculations for coronene with various aliphatic side chains demonstrate that the 3--4~$\mu$m spectrum depends sensitively on the type of aliphatic substitution \citep{2020ApJ...892...11B}.
Similarly, laboratory spectra of hydrogenated and methylated PAHs constrain the contribution of aliphatic groups to the 3.4~$\mu$m feature, indicating that different aliphatic modifications yield distinct spectral signatures \citep{2013ApJS..208...26S}.
The investigations of vinyl-substituted PAHs further reveal that such peripheral side groups alter the IR spectra relative to unsubstituted aromatic PAHs, and can introduce additional mid-IR sub-features in both neutral and cationic forms \citep{2012ApJ...755..120M, MAURYA20151}.
More recently, high-resolution experiments suggest that aromatic and aliphatic C--H vibrational modes can be strongly mixed within a single molecule, with additional contributions from combination and overtone bands \citep{2024MNRAS.535.3239E}. 
These results highlight the importance of including a wider range of molecular structures beyond purely aromatic PAHs in astronomical modeling.

In this study, we extend the established PAH diagnostic-grid framework by incorporating PAH molecules with aliphatic structural features from version 3.20 of the NASA Ames PAH IR Spectroscopic Database \citep{2010ApJS..189..341B,nasa2.00,nasa3.00ApJS23432B}. IR emission spectra for these species were generated following a methodology analogous to that employed by \citet{MPR2020,MPR2023}. By comparing the distribution of aliphatic-containing and purely aromatic PAHs within the diagnostic grid, we quantify how aliphatic components perturb commonly used band ratios and potentially bias the inferred PAH size and ionization state. Section \ref{sec:2} outlines the sample selection and characteristics; Section \ref{sec:3} describes the spectral generation and grid construction; Section \ref{sec:4} presents the aliphatic-induced grid perturbations and their implications; and Section \ref{sec:5} summarizes the main findings.

\section{Samples}\label{sec:2}

We compiled a dataset of 315 PAH molecules from Version 3.20 of the NASA Ames PAHdb database, which consists of 105 PAH molecules with aliphatic components and 210 purely aromatic PAH molecules. 
All molecular models employed in this work were the 
DFT-optimized structures retrieved directly from PAHdb, with no further DFT calculations conducted herein. It is well-established that smaller PAH molecules are susceptible to destruction by shocks, cosmic rays, and UV photons, making them unlikely to survive in the classical interstellar radiation field \citep{1996AA305602A,2020ApJ88817C,Tielens2011}. To investigate the influence of PAH molecules that can exist stably in typical interstellar environments on current diagnostic grids, we selected PAH species containing more than 20 carbon atoms.
For the sake of facilitating the analysis and ensuring consistency with the samples adopted in previous studies, we selected molecules composed exclusively of carbon and hydrogen atoms, excluding those containing heteroatoms such as nitrogen, oxygen, magnesium, silicon, or iron.
In addition, the selected sample molecules are required to contain isolated solo hydrogen atoms. The emission characteristics of 11.2 $\mu$m and 11.0 $\mu$m are attributed to the out-of-plane bending ($\text {CH}_{\text {oop}}$) modes of solo C–H bonds in neutral and cationic PAH molecules, respectively \citep{Tielens2008}. Therefore, to ensure the inclusion of molecules capable of producing these key diagnostic bands, we specifically selected PAH species containing solo C–H bonds for further analysis.

Sections \ref{sec:2.1} and \ref{sec:2.2} provide a comprehensive account of the selection procedures and the resulting characteristics of the aliphatic and aromatic PAH samples, respectively.
The unique identifiers (UIDs) of the PAHs used in this study, as listed in Version 3.20 of the NASA Ames PAHdb, are provided in Appendix \ref{uid list}.

\subsection{PAHs with aliphatic components}\label{sec:2.1}
\begin{figure}
	\includegraphics[width=\columnwidth]{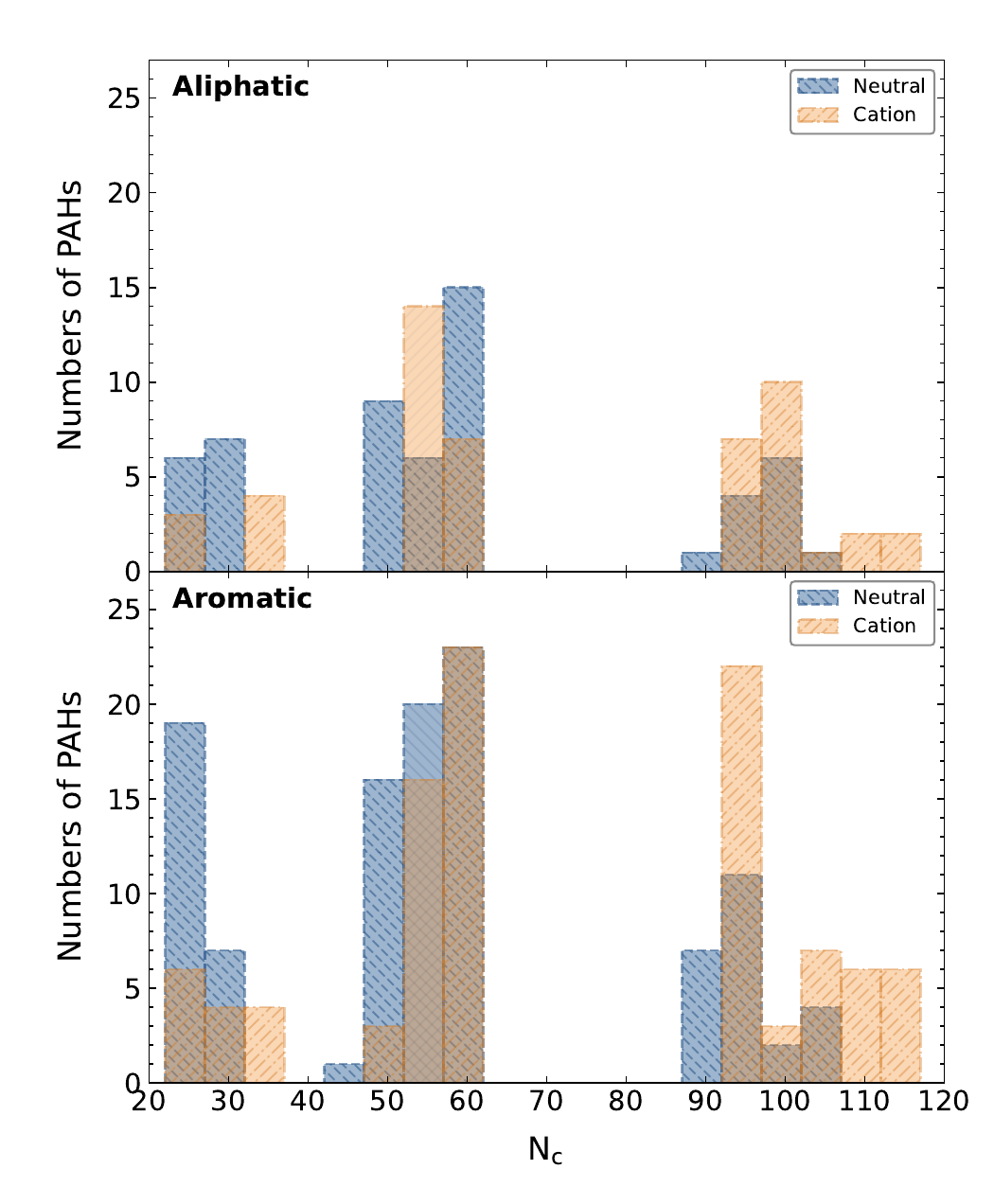}
    \caption{The upper panel shows the histograms of molecule counts for aliphatic PAH species, with blue representing neutral molecules and orange representing cations. The lower panel shows the corresponding histograms for purely aromatic PAH molecules, using the same color scheme.}
    \label{fig:histogram}
\end{figure}

In total, a sample of 55 neutral PAH molecules containing aliphatic components and 50 singly charged cationic PAH molecules with aliphatic components was compiled for this study. The neutral PAH molecules span a size range of 22 to 102 carbon atoms, with an average carbon atom $\langle N_{c} \rangle$ of 57. Among these, 37 molecules are superhydrogenated, while 18 contain methyl, methylene, or aliphatic carbon chain substituents. The cationic PAH molecules range from 22 to 112 carbon atoms, with an average carbon atom $ \langle N_{c} \rangle$ of 71. Of these, 22 are superhydrogenated and 28 possess methyl, methylene, or aliphatic carbon chains.
Although different types of aliphatic modifications may influence PAH band ratios in different ways
\citep{2012ApJ...755..120M,2013ApJS..208...26S,MAURYA20151,2020ApJ...892...11B}, the diagnostic grids in this work are constructed by mixing aliphatic-bearing and purely aromatic PAHs to generate samples spanning a prescribed aliphatic fraction. With Version 3.20 of the NASA Ames PAHdb, further subdividing the aliphatic-bearing sample into individual subtypes would reduce the available sample size to on the order of ten molecules per subtype in some cases, which is insufficient to robustly populate the grids and would lead to strong selection effects, particularly at high aliphatic fractions. Therefore, we focus on the overall influence of aliphatic content on the diagnostic grids. A more granular subtype-specific analysis will be feasible when larger and more uniformly sampled aliphatic PAH sets become available.
The resulting size distributions of the selected neutral and cationic aliphatic-bearing PAHs are shown in the upper panel of Figure~\ref{fig:histogram}.

It should be noted that in the PAHdb database, neutral aliphatic PAH molecules are noticeably absent in both carbon atom ranges of 30 to 45 and 61 to 85. Similarly, cationic aliphatic PAHs show missing data in the ranges of 35 to 50 and 61 to 95 carbon atoms. Based on these distributions, PAH molecules can be naturally classified into three size categories: small ($N_{c}$ $\leq$ 35), medium (45 < $N_{c}$ $\leq$ 60), and large ($N_{c}$ > 85).

\subsection{Pure aromatic PAHs}
\label{sec:2.2}
The NASA Ames PAHdb database contains 2,001 purely aromatic neutral PAH molecules and 493 purely aromatic cationic PAH molecules with more than 20 carbon atoms. To obtain our target aromatic PAH samples, we employed the k-nearest neighbors (KNN) algorithm, a widely used technique in machine learning, to perform sample selection.

The choice of this method is motivated mainly by two considerations. First, the molecular feature space in the PAHdb database comprises more than ten dimensions, incorporating both spectroscopic and structural parameters (e.g., atom count, molecular weight, edge structure descriptors, etc.). The KNN algorithm demonstrates strong robustness in handling such high-dimensional data sets with relatively small sample sizes (n < 3000). Its non-parametric nature also avoids the strong distributional assumptions typically required by conventional clustering methods \citep{hastie2009elements}. Second, our objective is to identify molecular populations that closely match the selected aliphatic PAHs. The Euclidean distance–based similarity metric employed by KNN enables a direct mapping of similarity thresholds for aliphatic molecules, offering better physical interpretability compared to black-box models such as support vector machines \citep{rudin2019stop}.

During molecular selection, we implemented a hierarchical weighting scheme for the set of parameters to mitigate interference from various confounding factors. Specifically: (1) Fundamental parameters (total weight: 10$\%$), including total atom number, molecular weight, total energy and zero point energy (each weighted at 2.5\%), were introduced to minimize biases arising from molecular size and basic physicochemical properties. (2) The parameters of the hydrogen edge structure (total weight: 40\%), which include five categories of edge configuration—solo hydrogens, duo hydrogens, trio hydrogens, quartet hydrogens, and quintet hydrogens (each weighted at 8\%) were adopted to suppress classification interference due to variations in edge structures. (3) The core parameter (total weight: 50\%) was explicitly centered on the carbon atom number of the PAH molecules, designed to emphasize the dominance of carbon skeleton characteristics in molecular screening by attenuating the influence of non-critical variables.

Finally, we identified a sample of purely aromatic PAH molecules that was twice the number of aliphatic PAHs to facilitate subsequent analyses. The neutral PAH molecules span a size range of 22 to 103 carbon atoms, with an average carbon atom $\langle N_{c} \rangle$ of 56 and the cationic PAH molecules range from 22 to 115 carbon atoms, with an average carbon atom $ \langle N_{c} \rangle$ of 72. Histograms showing the number distribution of purely aromatic neutral molecules and cations are presented in the lower panel of Figure~\ref{fig:histogram}.

\section{Methodology and Results}\label{sec:3}
To quantitatively investigate the influence of aliphatic PAHs on diagnostic grids, we constructed four sample groups characterized by varying ratios of aliphatic to aromatic molecules, specifically with aliphatic fractions of 0\% (purely aromatic), 33.3\%, 50\%, and 100\% (purely aliphatic). Each of these groups comprises spectra synthesized in five different ionization fractions, defined by the relative contributions of neutral and cationic species as follows: (1) Neutral (100\% neutral molecules), (2) N67C33 (67\% neutral and 33\% cationic), (3) N50C50 (50\% neutral and 50\% cationic), (4) N33C67 (33\% neutral and 67\% cationic) and (5) Cation (100\% cationic molecules). 
Neutral and cationic PAHs are treated as separate species in our analysis. We attempted to identify neutral–cation counterparts for the selected aliphatic-bearing PAHs. In our sample, 21 neutral–cation pairs correspond to the same molecular topology, while the remaining neutral and cationic species are not in a strict one-to-one correspondence due to the current PAHdb coverage and our selection filters. Therefore, the neutral and cationic grids are constructed from partially paired, partially independent molecular sets.
Section \ref{sec:3.1} provides a detailed description of the radiation model employed and the generation of characteristic spectra used in this study. Section \ref{sec:3.2} presents diagnostic grids corresponding to samples with varying aliphatic contents.

\subsection{Spectra and model}\label{sec:3.1}
Our PAH emission calculations employ the same excitation and emission framework as described in \citet{MPR2020, MPR2023}. This approach ensures direct comparability with the previously established diagnostic grids for purely aromatic PAHs, as the treatment of the radiation field and the band‑ratio measurement protocol remain identical. The goal of the present work is not to introduce modifications to the emission model itself, but to investigate how the incorporation of aliphatic‑bearing species affects the standard PAH size–charge diagnostics under controlled and internally consistent conditions. The primary extension in this study is the application of this well‑established framework to mixed aromatic–aliphatic PAH populations with defined aliphatic fractions. The specific assumptions and procedures used in this work are detailed below.

To generate PAH emission spectra from absorption spectra computed via DFT, we employed a radiation field model characterized by an average photon energy of 6 eV. The calculations incorporated the complete emission cascade of PAH molecules transitioning from the highest excited state to the vibrational ground state, comprehensively depicting the entire energy transfer pathway from molecular excitation to radiative relaxation \citep{nasa3.00ApJS23432B}. To enhance the generalizability of results across diverse observational targets, we did not utilize the classical Interstellar Radiation Field (ISRF) model \citep{1983}.
Such targets include reflection nebulae, H II regions, young stellar objects, asymptotic giant branch  stars, planetary nebulae, and galaxies, whose environments are not necessarily exposed to the classical interstellar radiation field \citep[see][]{MPR2020}.

Under typical astronomical observational conditions, PAH molecules might be in highly vibrationally excited states; consequently, their emission spectra inherently reflect anharmonic effects arising from molecular vibrational modes \citep{1998ApJ493793C,2002AA38639P}.  To account for and correct these anharmonicities in spectral analysis, \cite{nasa2.00} proposed applying a systematic redshift of approximately 15 $ \mathrm {cm}^{-1}$ to the computed emission band positions relative to their corresponding absorption bands. This adjustment ensures better alignment between theoretical predictions and observational data, and has been widely adopted in relevant studies. However,
\cite{10.1063/1.5038725} investigated 20 small PAHs with carbon atom numbers in the range of  $\mathrm {N_{c}}$ = 10--18, and demonstrated that the magnitude and direction of the pseudo-shifts induced by anharmonicity are dependent on both the molecular species and the characteristics of the vibrational bands. Furthermore, these pseudo-shifts typically remain below 15 $ \mathrm {cm}^{-1}$. \cite{MPR2020} also found that the relationships between the number of carbon atoms and PAH intensity ratios do not depend on the application of such systematic redshifts. 
We performed the same sensitivity test for our own sample (Appendix~\ref{sec:app_anharmonicity}) and likewise find that the main diagnostic-grid trends are insensitive to the application of a uniform 15~$\mathrm{cm}^{-1}$ redshift. Consequently, we did not apply any systematic redshift corrections to the spectra calculated in this study.

A separate issue relates to the influence of anharmonicity on the intrinsic band intensities, especially on the
$I_{11.2}/I_{3.3}$ intensity ratio, which is widely used as a PAH size diagnostic. As shown in Appendix~\ref{sec:app_anharmonicity}, such effects are more likely to alter the absolute calibration of the
$I_{11.2}/I_{3.3}$-based size diagnostic
rather than the main relative grid trends. 
Apart from the anharmonicity, the $I_{11.2}/I_{3.3}$ intensity ratio may also be modified by recurrent electronic fluorescence; furthermore, the intrinsic strength of the
 3.3 $\mu$m feature could have been overestimated by
quantum-chemical calculations \citep[see the recent review by][]{Tielens_2026}. 
We emphasize that our focus in this paper is to examine the trends in the band intensity ratio upon the introduction of aliphatic components, which should be independent of these effects.

The computed PAH spectral bands were convolved using a Lorentzian profile function with a full width at half-maximum (FWHM) of 15 $ \mathrm {cm}^{-1}$. While literature commonly employs various spectral fitting profiles (e.g., Lorentzian, Gaussian, and composite functions) and observed PAH emission bands in astronomical objects typically exhibit notable asymmetries and anharmonic characteristics \citep{1987ApJ315L61B, 2002AA38639P}, previous studies have shown that the choice of line profile has a negligible impact on PAH spectral observations \citep{2007ApJ655770S, 2008ApJ679310G, 2019ApJ871124S}. Thus, using a Lorentzian profile does not affect our conclusions about trends and relative intensity relationships among characteristic PAH emission bands.

 We measured the fluxes of the PAH emission bands at 3.3, 6.2, 7.7, 8.6, 11.0, and 11.2 $\mu$m, adopting integration wavelength ranges consistent with those defined by \citet{MPR2020}, and specific ranges are provided in Table \ref{tab:tab1}. Our measurement method closely follows that of \citet{MPR2020}, and plateau emission features were not treated separately. These plateau emissions typically exhibit distinct spatial distributions and are likely associated with larger PAH species \citep{2012ApJ74744P,2017ApJ836198P}. The same integration wavelength ranges were applied to neutral and cationic PAHs, particularly for the 11.0--11.5 $\mu$m region, which corresponds primarily to the out-of-plane C–H bending modes of solo hydrogens 
 \citep{2001A&A...370.1030H,2018ApJ...854..115R}. Generally, neutral PAHs dominate the 11.2 $\mu$m emission band, whereas cationic PAHs contribute predominantly to the 11.0 $\mu$m band. However, exact peak positions may change due to factors such as carbon atom number parity and molecular edge structures \citep{2008ApJ...678..316B,2009ApJ...697..311B,2018ApJ...854..115R}. 
 Consequently, following \citet{MPR2020}, we slightly broadened the integration ranges for the 11.0 and 11.2 $\mu$m bands and
 did not distinguish between these features in subsequent analysis, collectively referring to both as  the 11.2 $\mu$m band.
 
 \begin{table}
	\centering
	\caption{Integration Wavelength Ranges for PAH Emission Features.}
	\label{tab:tab1}
	\begin{tabular}{lcccccc} 
		\hline
		PAH feature & & & &  & Waveband (\si{\micro\meter})\\
		\hline
		3.3 \si{\micro\meter} &  & & & & 3.1--3.5\\
		6.2 \si{\micro\meter} &  & & & & 6.1--6.8\\
		7.7 \si{\micro\meter} &  & & & & 7.0--8.2\\
        8.6 \si{\micro\meter} &  & & & & 8.2--8.9\\
        11.2(11.0) \si{\micro\meter} & & & &  & 10.5--11.6\\
		\hline
	\end{tabular}

\end{table}

\begin{figure}
	\includegraphics[width=\columnwidth]{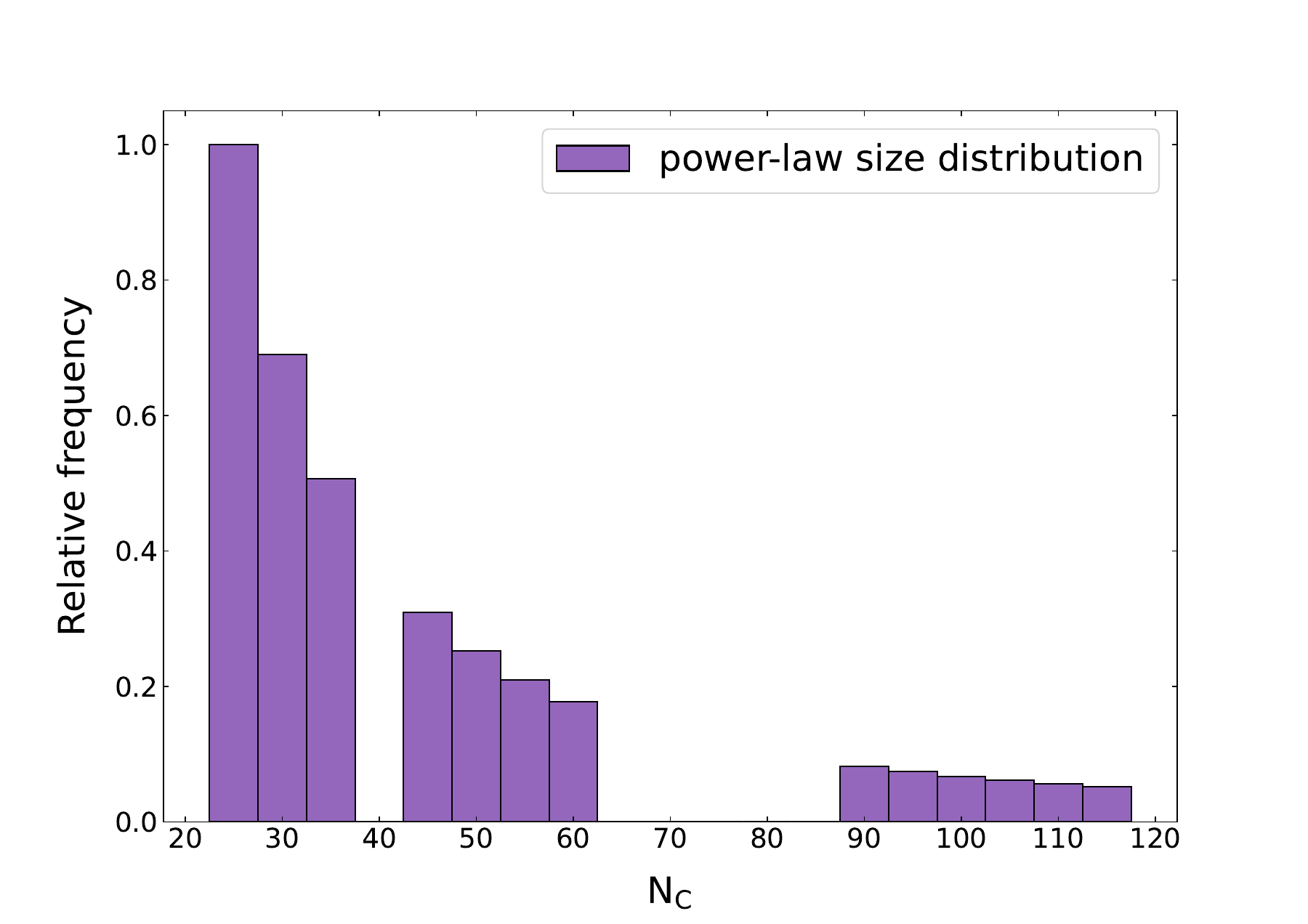}
    \caption{Distribution of carbon atom numbers in PAHs under a power-law distribution assumption, with frequencies normalized relative to that of the first bin.}
    \label{fig:power_law}
\end{figure}

Following the approach introduced by \cite{MPR2023}, we consider two distinct scenarios regarding the evolutionary processes of PAH populations: (1) smaller PAH molecules are preferentially destroyed (or equivalently, larger PAHs form more efficiently), a scenario we denote as SPR (``small PAHs removed''), resulting in an increase in the average molecular size of the PAH population; (2) larger PAH molecules are preferentially destroyed (or equivalently, smaller PAHs form more efficiently), referred to here as LPR (``large PAHs removed''), leading to a decrease in the average molecular size of the PAH population. To investigate the two evolutionary scenarios, we divided the PAH molecules into 19 bins based on their carbon atom number, with each bin spanning a width of $\Delta {N_{c}}$ = 5 and covering the range from ${N_{c}}$ = 22 to  115. Each bin is represented by the central value of its carbon-atom-number interval. As mentioned in Section \ref{sec:2.1}, certain carbon-atom-number intervals lacked PAH data; consequently, the final analysis included 13 bins with available molecular data. For each bin, we combined and averaged the spectral data of the included PAH molecules to derive a representative spectrum, which was subsequently weighted according to the power-law size distribution described by \cite{1993ApJ...415..397S}: $\mathrm{d}N_{\mathrm{PAH}} / N_{\mathrm{H}} = B_{\mathrm{c}}N_{\mathrm{c}}^{-\beta - 1}\mathrm{d}N_{\mathrm{c}}$, where $\mathrm{d}N_{\mathrm{PAH}} / N_{\mathrm{H}}$ denotes the number of PAH molecules per interstellar hydrogen atom within the carbon number range $N_{\mathrm{c}}$ and $N_{\mathrm{c}} + dN_{\mathrm{c}}$. Following \cite{1993ApJ...415..397S} and \cite{MPR2023}, we adopted a normalization parameter $B_{\mathrm{c}} = 1.24\times 10^{-6}$ and a power law exponent $\beta =0.833$, with the assumption of an interstellar hydrogen column density $N_{\text{H}} = 1.9\times 10^{21}\,\text{cm}^{-2}$. Figure \ref{fig:power_law} shows the power law distribution of the samples.

\begin{figure}
	\includegraphics[width=\columnwidth]{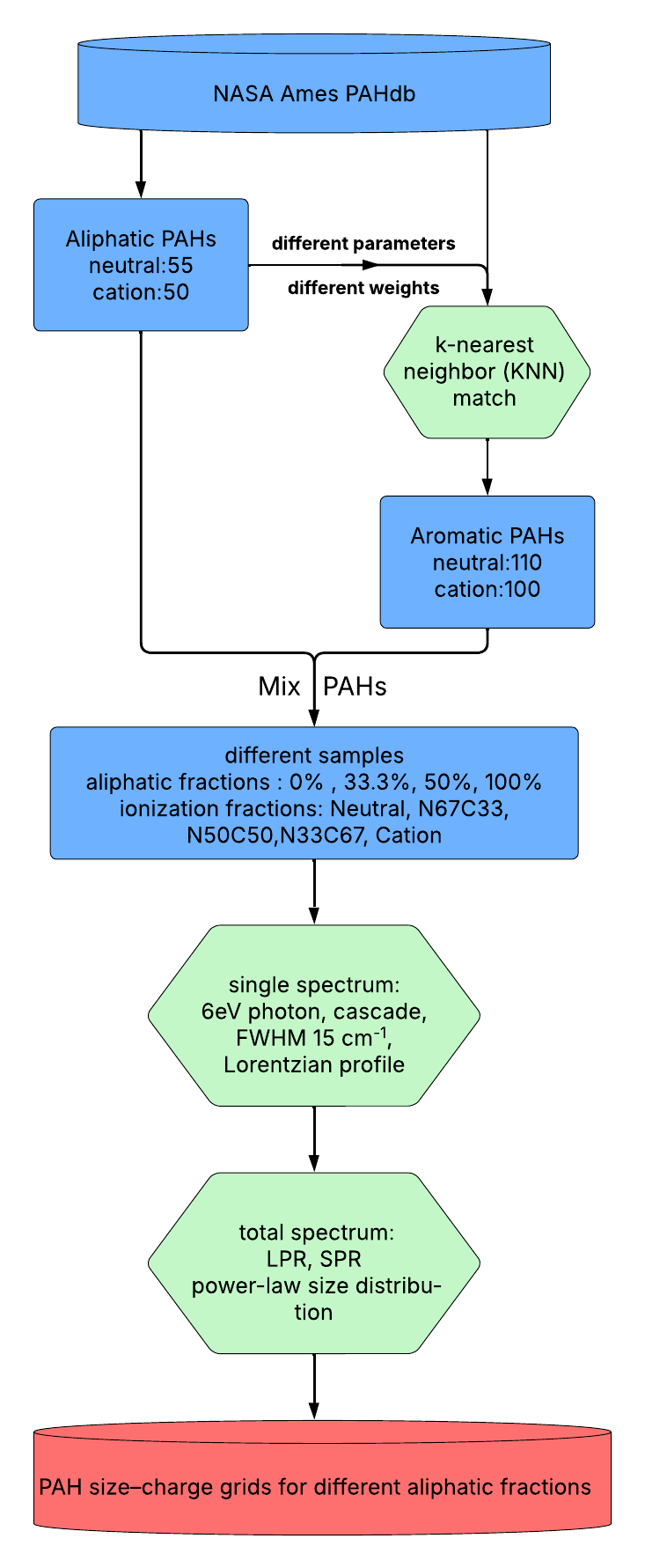}
    \caption{Flowchart for constructing the diagnostic grid of
     PAH size and ionization fractions.}
    \label{fig:flowchart}
\end{figure}

To reproduce the previously defined SPR and LPR scenarios, we employed a stepwise subtraction method on the total spectrum. To simplify calculations, the sample was divided into three size categories: small, medium, and large, based on the distribution described in Section \ref{sec:2.1}. For the SPR scenario, we sequentially subtracted contributions from the small and medium size bins from the total spectrum; in contrast, for the LPR scenario, contributions from the large and medium size bins were successively removed. Following these procedures, we obtained the final weighted average spectra for both scenarios, which were used for subsequent comparative analysis of PAH spectral evolution. An overview of the workflow, from sample selection and spectral generation to diagnostic grid construction, is provided in Figure \ref{fig:flowchart}.

\subsection{PAH size–charge grids for samples containing aliphatic components}\label{sec:3.2}
\begin{figure*}
	\includegraphics[width=\textwidth]{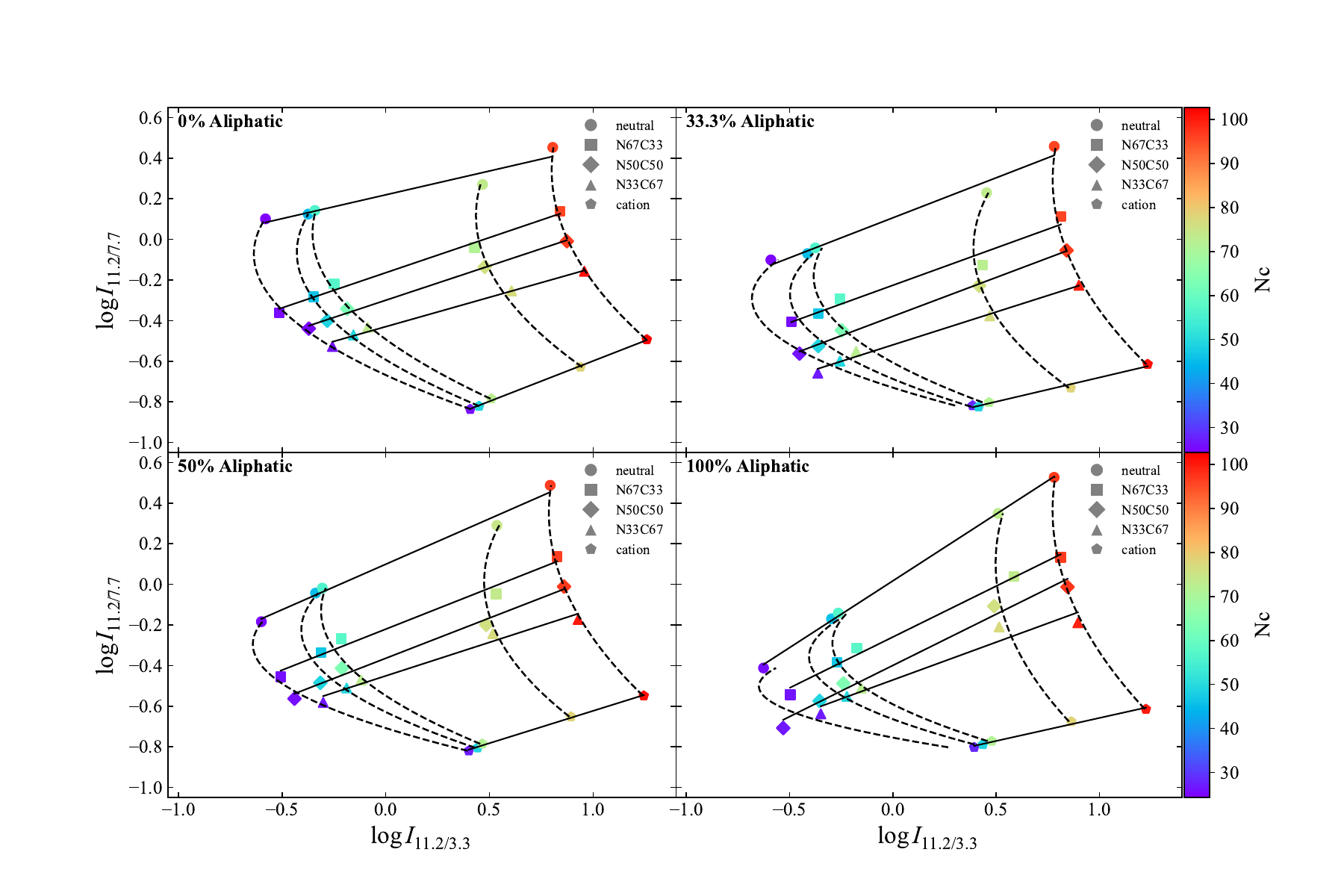}
    \caption{PAH size–ionization diagnostic grids for samples with varying aliphatic contents of 0$\%$, 33.3$\%$, 50$\%$, and 100$\%$.  Data points represent mixtures of PAH species, color-coded by their number of carbon atoms $\mathrm {N_{c}}$, as indicated by the adjacent color bar. Parameters of each point are listed in Table \ref{tab:tab2}. Linear fits connect points with the same ionization fraction; corresponding fit parameters are provided in Table \ref{tab:tab3}. Parabolic fits connect points with similar molecular sizes across different ionization states; the parameters of these fits are listed in Table \ref{tab:tab4}. 
    }
    \label{fig:fig4}
\end{figure*}

To investigate the influence of aliphatic components on the determination of charge states and size distributions of astronomical PAHs, we constructed diagnostic diagrams using samples with varying aliphatic contents, and superimposed diagnostic grids onto these diagrams.
Figure \ref{fig:fig4} presents a set of PAH size–ionization diagnostic grids, each corresponding to a model PAH population with a different fraction of aliphatic content (0$\%$, 33.3$\%$, 50$\%$, and 100$\%$ aliphatic sample). In each panel, the horizontal axis plots the intensity ratio $I_{11.2/3.3}$, while the vertical axis plots $I_{11.2/7.7}$. These particular band ratios serve as diagnostic tracers. The intensity ratio $I_{11.2/3.3}$ increases with PAH molecular size, as characterized by $N_{\rm C}$ \citep{MPR2020}, while $I_{11.2/7.7}$ decreases as PAHs become more ionized \citep{2021MNRAS.504.5287R}. Consequently, movement to the right on the diagnostic grid indicates larger PAH sizes, whereas upward movement corresponds to a more neutral, less ionized PAH population. Each data point in a panel represents a PAH population characterized by an average number of carbon atoms and a specific charge state. The points are color-coded by $N_{\rm C}$, as indicated by the color bar adjacent to each grid. The exact parameters for each data point, including the size and  ionization fraction
of the PAHs, are listed in Table \ref{tab:tab2}. 
\begin{table*}
\centering
\caption{Parameters of PAH Samples Used in the Diagnostic Grids.}
\label{tab:tab2} 
\resizebox{\textwidth}{!}{ 
\begin{tabular}{cccccccccccccccccccccccc}
\hline
Sample&\multicolumn{3}{c}{Neutral} && \multicolumn{3}{c}{N67C33} && \multicolumn{3}{c}{N50C50} &&\multicolumn{3}{c}{N33C67} && \multicolumn{3}{c}{Cation}\\
& $\mathrm {N_{c}}$ & $\log\frac{11.2}{7.7}$ & $ \log\frac{11.2}{3.3}$ & &$\mathrm {N_{c}}$ & $\log\frac{11.2}{7.7}$ & $\log\frac{11.2}{3.3}$ &&$\mathrm {N_{c}}$ & $\log\frac{11.2}{7.7}$ & $\log\frac{11.2}{3.3}$ &&$\mathrm {N_{c}}$ & $\log\frac{11.2}{7.7}$ & $\log\frac{11.2}{3.3}$ &&$\mathrm {N_{c}}$ & $\log\frac{11.2}{7.7}$ & $\log\frac{11.2}{3.3}$\\
\hline
0$\%$	&	24 	&	0.10 	&	-0.58 	&&	25 	&	-0.36 	&	-0.51 	&&	25 	&	-0.44 	&	-0.37 	&&	26 	&	-0.53 	&	-0.26 	&&	26 	&	-0.84 	&	0.41 	\\
Aliphatic	&	45 	&	0.12 	&	-0.37 	&&	46 	&	-0.28 	&	-0.35 	&&	49 	&	-0.40 	&	-0.28 	&&	49 	&	-0.47 	&	-0.16 	&&	49 	&	-0.82 	&	0.45 	\\
	&	56 	&	0.14 	&	-0.34 	&&	57 	&	-0.22 	&	-0.25 	&&	63 	&	-0.34 	&	-0.19 	&&	73 	&	-0.43 	&	-0.09 	&&	73 	&	-0.79 	&	0.51 	\\
	&	74 	&	0.27 	&	0.47 	&&	71 	&	-0.04 	&	0.43 	&&	76 	&	-0.13 	&	0.48 	&&	77 	&	-0.25 	&	0.61 	&&	78 	&	-0.63 	&	0.94 	\\
	&	96 	&	0.45 	&	0.81 	&&	96 	&	0.14 	&	0.84 	&&	97 	&	-0.01 	&	0.87 	&&	101 	&	-0.16 	&	0.96 	&&	103 	&	-0.49 	&	1.26 	\\
    \\
33.3$\%$	&	25 	&	-0.10 	&	-0.59 	&&	26 	&	-0.41 	&	-0.49 	&&	26 	&	-0.56 	&	-0.45 	&&	27 	&	-0.66 	&	-0.36 	&&	28 	&	-0.82 	&	0.39 	\\
Aliphatic	&	46 	&	-0.07 	&	-0.41 	&&	47 	&	-0.37 	&	-0.36 	&&	49 	&	-0.53 	&	-0.36 	&&	50 	&	-0.60 	&	-0.26 	&&	49 	&	-0.82 	&	0.41 	\\
	&	56 	&	-0.04 	&	-0.38 	&&	58 	&	-0.29 	&	-0.26 	&&	64 	&	-0.45 	&	-0.25 	&&	72 	&	-0.55 	&	-0.18 	&&	72 	&	-0.80 	&	0.47 	\\
	&	74 	&	0.23 	&	0.45 	&&	73 	&	-0.13 	&	0.43 	&&	76 	&	-0.23 	&	0.42 	&&	77 	&	-0.38 	&	0.47 	&&	78 	&	-0.73 	&	0.86 	\\
	&	96 	&	0.46 	&	0.78 	&&	96 	&	0.11 	&	0.81 	&&	97 	&	-0.05 	&	0.84 	&&	100 	&	-0.23 	&	0.90 	&&	101 	&	-0.62 	&	1.23 	\\
    \\
50$\%$	&	25 	&	-0.18 	&	-0.60 	&&	26 	&	-0.45 	&	-0.51 	&&	26 	&	-0.56 	&	-0.44 	&&	27 	&	-0.58 	&	-0.30 	&&	27 	&	-0.82 	&	0.40 	\\
Aliphatic	&	45 	&	-0.04 	&	-0.34 	&&	46 	&	-0.34 	&	-0.31 	&&	49 	&	-0.49 	&	-0.32 	&&	49 	&	-0.51 	&	-0.19 	&&	49 	&	-0.81 	&	0.44 	\\
	&	56 	&	-0.02 	&	-0.31 	&&	58 	&	-0.27 	&	-0.21 	&&	63 	&	-0.41 	&	-0.21 	&&	73 	&	-0.47 	&	-0.12 	&&	72 	&	-0.79 	&	0.47 	\\
	&	75 	&	0.29 	&	0.54 	&&	74 	&	-0.05 	&	0.53 	&&	76 	&	-0.20 	&	0.48 	&&	77 	&	-0.24 	&	0.52 	&&	78 	&	-0.65 	&	0.89 	\\
	&	97 	&	0.49 	&	0.79 	&&	96 	&	0.14 	&	0.83 	&&	97 	&	-0.01 	&	0.86 	&&	100 	&	-0.17 	&	0.93 	&&	101 	&	-0.55 	&	1.24 	\\
    \\
100$\%$	&	26 	&	-0.41 	&	-0.63 	&&	26 	&	-0.54 	&	-0.50 	&&	27 	&	-0.71 	&	-0.53 	&&	27 	&	-0.64 	&	-0.35 	&&	28 	&	-0.80 	&	0.39 	\\
Aliphatic	&	46 	&	-0.17 	&	-0.30 	&&	47 	&	-0.38 	&	-0.27 	&&	49 	&	-0.57 	&	-0.36 	&&	50 	&	-0.55 	&	-0.22 	&&	49 	&	-0.79 	&	0.44 	\\
	&	57 	&	-0.14 	&	-0.26 	&&	58 	&	-0.31 	&	-0.18 	&&	64 	&	-0.49 	&	-0.24 	&&	72 	&	-0.51 	&	-0.15 	&&	71 	&	-0.77 	&	0.48 	\\
	&	73 	&	0.35 	&	0.51 	&&	74 	&	0.04 	&	0.59 	&&	76 	&	-0.11 	&	0.49 	&&	77 	&	-0.21 	&	0.51 	&&	78 	&	-0.68 	&	0.86 	\\
	&	97 	&	0.53 	&	0.78 	&&	97 	&	0.13 	&	0.81 	&&	97 	&	-0.01 	&	0.85 	&&	99 	&	-0.19 	&	0.90 	&&	100 	&	-0.61 	&	1.22 	\\
\hline
\end{tabular}
}
\end{table*}
Within each panel of Figure \ref{fig:fig4}, lines are drawn to connect data points, emphasizing systematic trends. First, data points sharing the same PAH ionization fraction (defined as the percentage of PAHs in the ionized state) are linked by straight-line segments. These nearly linear sequences illustrate how band ratios vary with PAH size under a fixed ionization fraction. 
Each sequence is fitted with a linear regression using the equation:
\begin{equation}
    \log I_{11.2/7.7} = k \log I_{11.2/3.3} + b.
\end{equation}
The slopes ($k$) and intercepts ($b$) of these linear fits are presented in Table \ref{tab:tab3}.
\begin{table}
	\caption{Linear Fit Parameters for Constant Ionization Sequences.}
	\label{tab:tab3}
	\begin{tabular}{lccccc} 
		\hline
		Sample & Charge state & $k\pm \Delta k$ & $b\pm \Delta b$ & r  \\
		\hline
0$\%$	&	Neutral	&	0.24 	±	0.04 	&	0.22 	±	0.02 	&	0.966 	\\
	&	N67C33	&	0.35 	±	0.02 	&	-0.16 	±	0.01 	&	0.994 	\\
	&	N50C50	&	0.34 	±	0.01 	&	-0.30 	±	0.01 	&	0.997 	\\
	&	N33C67	&	0.29 	±	0.02 	&	-0.43 	±	0.01 	&	0.994 	\\
	&	Cation	&	0.40 	±	0.01 	&	-1.00 	±	0.01 	&	0.999 	\\
    \\
33.3$\%$	&	Neutral	&	0.39 	±	0.04 	&	0.11 	±	0.02 	&	0.987 	\\
	&	N67C33	&	0.37 	±	0.04 	&	-0.23 	±	0.02 	&	0.983 	\\
	&	N50C50	&	0.38 	±	0.02 	&	-0.38 	±	0.01 	&	0.998 	\\
	&	N33C67	&	0.32 	±	0.02 	&	-0.52 	±	0.01 	&	0.995 	\\
	&	Cation	&	0.24 	±	0.02 	&	-0.92 	±	0.01 	&	0.991 	\\
    \\
50$\%$	&	Neutral	&	0.45 	±	0.03 	&	0.10 	±	0.02 	&	0.993 	\\
	&	N67C33	&	0.40 	±	0.03 	&	-0.22 	±	0.02 	&	0.989 	\\
	&	N50C50	&	0.40 	±	0.03 	&	-0.36 	±	0.01 	&	0.993 	\\
	&	N33C67	&	0.33 	±	0.03 	&	-0.45 	±	0.02 	&	0.986 	\\
	&	Cation	&	0.32 	±	0.01 	&	-0.94 	±	0.01 	&	0.999 	\\
    \\
100$\%$	&	Neutral	&	0.66 	±	0.01 	&	0.02 	±	0.01 	&	0.999 	\\
	&	N67C33	&	0.50 	±	0.03 	&	-0.26 	±	0.01 	&	0.996 	\\
	&	N50C50	&	0.51 	±	0.04 	&	-0.40 	±	0.02 	&	0.991 	\\
	&	N33C67	&	0.37 	±	0.05 	&	-0.47 	±	0.03 	&	0.973 	\\
	&	Cation	&	0.23 	±	0.01 	&	-0.89 	±	0.01 	&	0.994 	\\

		\hline
	\end{tabular}\\
\footnotesize{Note: Slopes and intercepts of linear fits connecting PAH samples with the same ionization fraction in the diagnostic grids. Fits follow the form $\log I_{11.2/7.7} = k \log I_{11.2/3.3} + b$. $r$ denotes the correlation coefficient.}
\end{table}
Second, data points corresponding to similar PAH sizes (i.e., similar
$N_{\rm c}$) are connected across different charge states. These trajectories, for instance, linking a neutral PAH to its cationic counterparts of the same size, exhibit distinct curvature. To capture the non-linear variation in band ratios as molecules transition from neutral to ionized forms, we fit each size-specific sequence with a parabolic function using the equation:
\begin{equation}
\label{avalue}
\log I_{11.2/3.3} = A \log I_{11.2/7.7}^{2} + B \log I_{11.2/7.7} + C.
\end{equation}
The coefficients of these parabolic fits are provided in Table \ref{tab:tab4}.
\begin{table}
	\caption{Parabolic Fit Parameters for Constant Size Sequences.}
	\label{tab:tab4}
	\begin{tabular}{lcccccc} 
		\hline
		Sample & $N_{c}$ & $A\pm \Delta A$ & $B\pm \Delta B$ & $C\pm \Delta C$ \\
		\hline
0$\%$	&	26	&	1.79 	±	0.12 	&	-6.31 	±	0.35 	&	6.52	±	0.25 	\\
	&	48	&	1.53 	±	0.10 	&	-5.39 	±	0.31 	&	5.68 	±	0.22 	\\
	&	64	&	1.29 	±	0.09 	&	-4.68 	±	0.27 	&	5.24 	±	0.20 	\\
	&	76	&	0.92 	±	0.20 	&	-3.21 	±	0.57 	&	4.67 	±	0.40 	\\
	&	98	&	0.66 	±	0.05 	&	-2.41 	±	0.15 	&	4.81 	±	0.11 	\\
\\															
33.3$\%$	&	26	&	3.53 	±	1.35 	&	-10.80 	±	3.70 	&	9.15 	±	2.47 	\\
	&	48	&	2.84 	±	0.72 	&	-8.83 	±	2.02 	&	7.79 	±	1.37 	\\
	&	64	&	2.12 	±	0.68 	&	-6.86 	±	1.89 	&	6.56 	±	1.28 	\\
	&	76	&	0.97 	±	0.14 	&	-3.29 	±	0.43 	&	4.58 	±	0.31 	\\
	&	98	&	0.55 	±	0.08 	&	-2.10 	±	0.24 	&	4.55 	±	0.18 	\\
\\															
50$\%$	&	26	&	3.76 	±	0.57 	&	-11.50 	±	1.52 	&	9.70 	±	0.99 	\\
	&	48	&	2.49 	±	0.48 	&	-7.86 	±	1.34 	&	7.15 	±	0.91 	\\
	&	64	&	1.79 	±	0.37 	&	-5.92 	±	1.03 	&	5.89 	±	0.70 	\\
	&	76	&	0.93 	±	0.20 	&	-3.11 	±	0.59 	&	4.54 	±	0.43 	\\
	&	98	&	0.56 	±	0.05 	&	-2.13 	±	0.15 	&	4.60 	±	0.11 	\\
\\															
100$\%$	&	26	&	10.10 	±	7.35 	&	-26.40 	±	17.60 	&	18.10 	±	10.50 	\\
	&	48	&	4.29 	±	1.55 	&	-12.30 	±	4.09 	&	9.74 	±	2.65 	\\
	&	64	&	3.22 	±	1.02 	&	-9.56 	±	2.69 	&	8.06 	±	1.75 	\\
	&	76	&	0.58 	±	0.27 	&	-2.08 	±	0.82 	&	3.84 	±	0.60 	\\
	&	98	&	0.47 	±	0.06 	&	-1.85 	±	0.17 	&	4.38 	±	0.13 	\\

		\hline
	\end{tabular}\\
\footnotesize{Note: Coefficients of parabolic fits connecting PAH samples with similar sizes across varying ionization states. Fits follow the form $\log I_{11.2/3.3} = A \log I_{11.2/7.7}^{2} + B \log I_{11.2/7.7} + C$.}
\end{table}

As shown in  Figure \ref{fig:fig4}, linear and parabolic connections collectively form a grid in intensity-ratio space, serving as a visual diagnostic tool. Observed PAH band ratios can be mapped onto this grid to infer the characteristic PAH size and ionization state of astronomical sources. 
Note that each grid point corresponds to the intensity ratios obtained from a population‑averaged mixed PAH spectrum, computed for given aliphatic and ionization fractions, rather than from an individual PAH molecule.
A comparison of the four panels reveals that variations in aliphatic fractions induce  pronounced changes in the intensity ratio $I_{11.2}/I_{7.7}$,
whereas the corresponding changes in the $I_{11.2}/I_{3.3}$ intensity ratios are rather subtle.
As an illustration, consider a source with measured values of $\log(I_{11.2/3.3})=-0.2$  (i.e., dominated by small PAHs) and $\log(I_{11.2/7.7})=-0.4$. 
In the purely aromatic (0\% aliphatic) diagnostic grid, 
 this measurement corresponds to a PAH population with an ionization fraction of roughly 50\% and a size of about $N_{\rm C}\approx 56 \pm 19$.
If the same observational point is instead matched against a grid that includes aliphatic components, for example the 33.3\% aliphatic grid, the inferred ionization fraction falls below 50\%. This indicates a more neutral PAH population, while the inferred PAH size shifts to $N_{\rm C}\approx 65 \pm 17$ (i.e., dominated by medium‑size PAHs).
This reflects the impact of aliphatic content on the bonding strength and molecular configuration of PAHs.
The overall grid structure, including linear trends for constant ionization fractions and parabolic curves for constant sizes, remains largely intact across all panels.

\begin{figure*}
	\includegraphics[width=\textwidth]{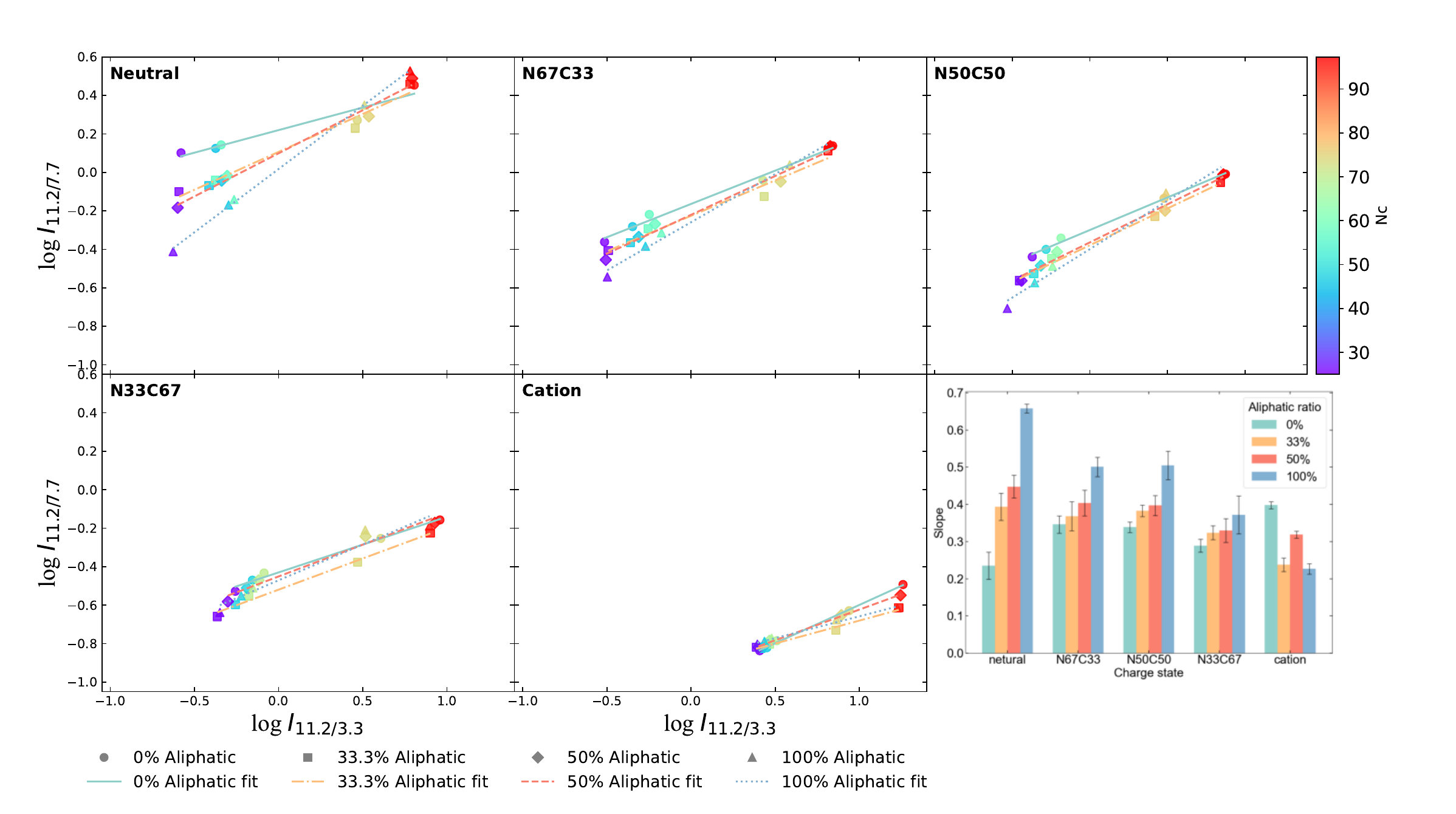}
    \caption{Diagnostic band ratio diagrams extracted from the PAH diagnostic grids (Figure \ref{fig:fig4}), illustrating the influence of varying aliphatic content (0$\%$, 33.3$\%$, 50$\%$, and 100$\%$) at five fixed ionization fractions (Neutral, N67C33, N50C50, N33C67, and Cation). Each panel presents data points representing individual PAH species or mixtures color-coded by their number of carbon atoms ($\mathrm {N_{c}}$). Linear fits (solid and dashed lines) connect points within each ionization fraction to quantify trends in diagnostic ratios with increasing aliphatic content. The bottom-right panel summarizes these trends by showing the slope distribution histogram of the linear fits across different ionization states, highlighting the systematic increase in slope with aliphatic fraction for all cases except the fully ionized  state.}
    \label{fig:fig5}
\end{figure*}

\section{Discussions}\label{sec:4}
\subsection{Impact of aliphatic components on the PAH ionization diagnostic}\label{sec:4.1}
It has long been recognized that certain mid-IR PAH band ratios, such as  $I_{11.2}/I_{7.7}$, serve as tracers of the PAH ionization state. However, the present results demonstrate that these diagnostic ratios can be markedly influenced by the molecular structure of PAHs, particularly by the presence of aliphatic components.
Figure~\ref{fig:fig5} illustrates how increasing aliphatic carbon fractions systematically affect key PAH band intensity ratios across different ionization states. Each horizontal series in Figure~\ref{fig:fig5} corresponds to a fixed PAH ionization fraction (from neutral up to fully ionized), with data points along a series representing varying PAH sizes. 
This figure is extracted from Figure~\ref{fig:fig4} by taking horizontal slices at fixed ionization fractions. Specifically, for each ionization case, we trace how the grid position changes as the aliphatic fraction increases from 0\% to 33.3\%, 50\%, and 100\%. In this manner, Figure~\ref{fig:fig5} highlights the impact of aliphatic content under a fixed charge-state mixture.
As a concrete example, we consider the N50C50 sequence. In the purely aromatic grid, the point at $N_{\mathrm C}\approx 63$ corresponds to $\log(I_{11.2}/I_{3.3})\approx -0.19$ and $\log(I_{11.2}/I_{7.7})\approx -0.34$. 
For a nearly identical characteristic size in the
 100\% aliphatic grid, the corresponding point shifts to $\log(I_{11.2}/I_{3.3})\approx -0.24$ and $\log(I_{11.2}/I_{7.7})\approx -0.49$. 
 Thus, even with the nominal ionization fraction held fixed, increasing aliphatic content shifts the point primarily downward in the diagnostic plane, i.e. towards a more ``cation-like'' location if interpreted using aromatic-only grids. 
 This example demonstrates that neglecting aliphatic structure can bias ionization estimates, even when the underlying charge‑state mixture remains unchanged.

A distinct trend emerges from the data for neutral and partially ionized mixtures with a low proportion of ionized PAHs. Specifically, the slope of the linear fit in the diagnostic ratio plane 
($I_{11.2}/I_{7.7}$ versus $I_{11.2}/I_{3.3}$) 
steepens as the aliphatic content increases.
Notably, the effect of increasing aliphatic content on the diagnostic ratio and fitting slope is pronounced for $N_{\rm C} \lesssim 60$.
For this size regime, data points shift toward lower
$I_{11.2}/I_{7.7}$ values as the aliphatic fraction increases from  0$\%$ to 100$\%$. 
In contrast, this behavioral pattern becomes less pronounced for larger $N_{\rm C}$.
For mixtures with a high proportion of ionized PAHs, however, the diagnostic slope remains nearly unchanged or even flattens as the aliphatic fraction increases
(Figure~\ref{fig:fig5}).
This finding underscores that PAH charge-state indicators are sensitive to molecular structure:  shifts in $I_{11.2}/I_{7.7}$ can stem 
not only from PAH charge state, as traditionally thought, but also from aliphatic-related structural modifications. Therefore, PAH ionization diagnostics may be subject to significant bias when aliphatic effects are overlooked.

This trend reflects how PAH emission signatures are tuned by the integration of aliphatic components into the aromatic framework.
Specifically, in-plane C–H bending modes are dominated by the electronic perturbations induced by aliphatic side groups: electron-donating substituents increase the force constant of in-plane bending vibrations and thereby elevate their IR emission intensities. In contrast, CH$_{\rm oop}$ modes are highly susceptible to steric distortions. Bulky aliphatic side chains introduce steric hindrance that distorts the planar aromatic skeleton, while superhydrogenation further induces non-planarity by incorporating excess hydrogen atoms onto ring carbon sites; both effects reduce the amplitude of CH$_{\rm oop}$ motions and attenuate their corresponding IR emission intensities. 
Additionally, when an aliphatic side group substitutes for a solo aromatic H atom, the corresponding solo $\mathrm{CH_{oop}}$ vibrational mode is eliminated if this is the only solo site on the PAH molecule, or reduced in total intensity if multiple independent solo sites are present. This substitution also produces aliphatic infrared features, most notably the 7.25\,$\mu$m band associated with the asymmetric scissoring mode of methyl  groups.
As a result, the diagnostic ratio $I_{11.2}/I_{7.7}$ exhibits a monotonic decrease with increasing aliphatic fraction.
For a given PAH species, the impact of aliphatic moiety incorporation intensifies with an increasing aliphatic-to-aromatic ratio; consequently, this effect is less pronounced for PAHs bearing extended aromatic planes.
Furthermore, the delocalization of positive charge redistributes electron density throughout the PAH framework, thereby reducing the polarity discrepancy between aromatic and aliphatic C–H bonds. This accounts for the negligible dependence of the intensity ratios $I_{11.2}/I_{7.7}$ versus
$I_{11.2}/I_{3.3}$ on the aliphatic fraction in fully ionized PAHs.

As reported by \citet{2025A&A...699A.133K} using state-of-the-art JWST observational data, the 11.2\,$\mu$m band exhibits a distinct bimodal profile, with a secondary component emerging on its 
long-wavelength wing, in the more shielded regions of the Orion Bar photodissociation region (PDR). These regions are  characterized by prominent aliphatic emission bands. A plausible interpretation of this phenomenon invokes the presence of superhydrogenated PAHs: extensive hydrogenation induces structural saturation in these molecules, shifting their characteristic emission toward higher frequencies while drastically suppressing their emission intensity. This behavior prevents them from obscuring the faint long-wavelength component of the 
11.2\,$\mu$m band. 
This spectral trend supports the notion that aliphatic components can influence the diagnostic performance of the $I_{11.2}/I_{7.7}$ ratio.

In summary, interpretations of PAH band ratios must account for structural biases. Astronomical diagnostics relying solely on aromatic PAH templates risk misestimating the true PAH charge state in the presence of aliphatic substituents. As shown, aliphatic substitution modulates the 11.2\,$\mu$m band relative to the 7.7\,$\mu$m
complex in a size- and ionization-dependent manner. Applying diagnostic grids calibrated on purely aromatic PAHs to aliphatic-containing observations will systematically bias inferred PAH charge and size distributions.
Our results highlight that aliphatic content is a critical third parameter, alongside size and charge, governing PAH emission ratios. Caution is therefore warranted when attributing band ratio variations exclusively to ionization state changes. Environments fostering aliphatic-rich PAHs (e.g., shielded regions around cool stars) may yield spectra erroneously interpreted as indicative of excessive ionization 
under aromatic-only assumptions. Recognizing structural diversity, especially aromatic-aliphatic mixtures, will enable more accurate PAH charge state assessments across diverse cosmic environments.

\subsection{Impact of aliphatic components on the PAH size diagnostic}\label{sec:4.2}

\begin{figure*}
	\includegraphics[width=\textwidth]{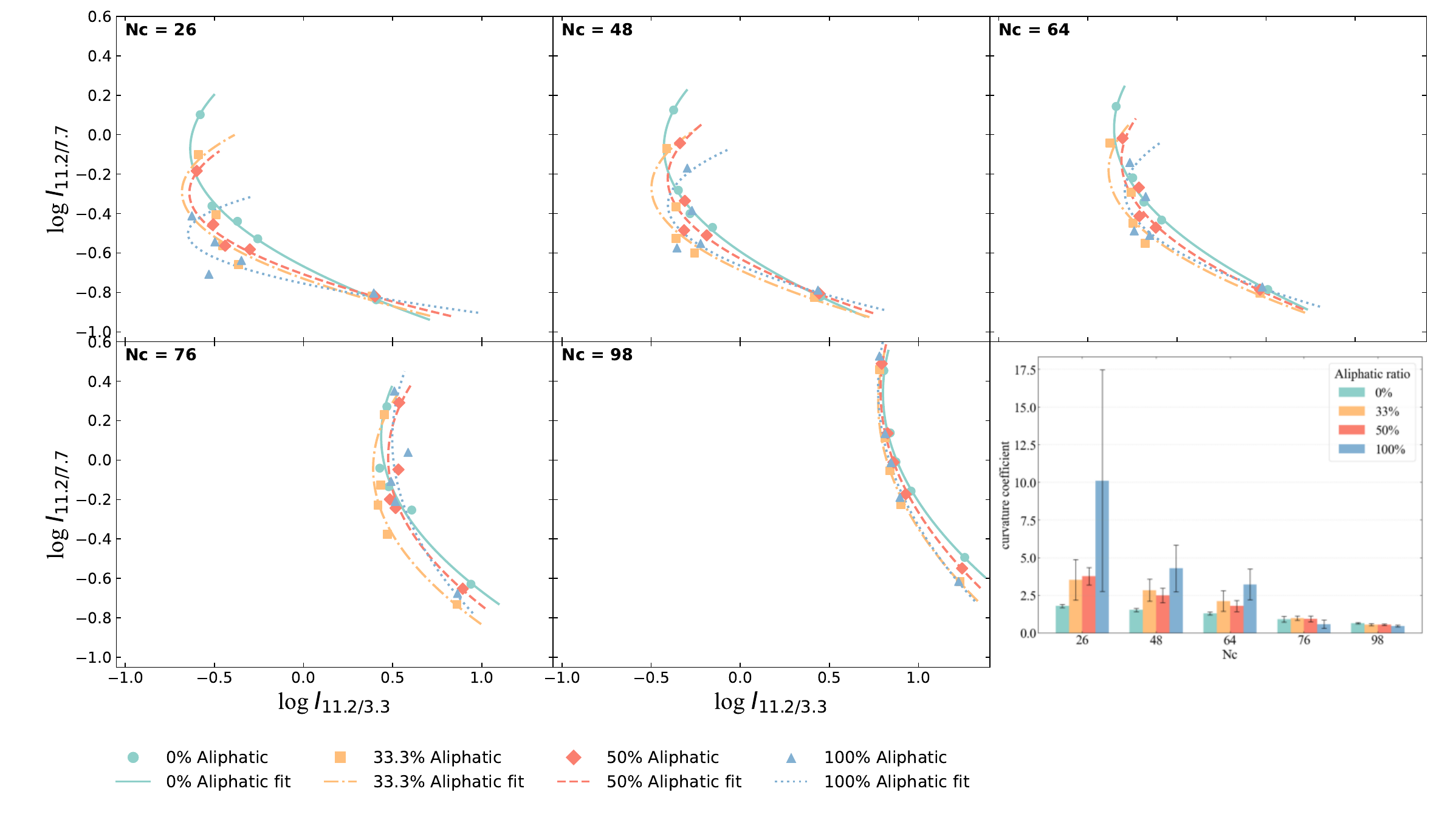}
    \caption{Diagnostic sequences illustrating the impact of aliphatic fraction on PAH diagnostic ratios at fixed carbon numbers. Each sub-panel displays parabolic fits connecting points of identical $N_{\rm C}$ extracted from diagnostic grids with different aliphatic content (0$\%$, 33.3$\%$, 50$\%$, and 100$\%$), originally shown in Figure 4. The curvature of these sequences quantifies the sensitivity of spectral ratios to aliphatic substitution. The bottom-right histogram summarizes the curvature coefficients  obtained from the parabolic fits, clearly demonstrating that smaller PAHs (lower $N_{\rm C}$) exhibit a significantly stronger sensitivity to changes in aliphatic content compared to larger PAHs.}
    \label{fig:fig6}  
\end{figure*}

As illustrated in Figure~\ref{fig:fig6}, the intensity ratio $I_{11.2}/I_{3.3}$ acts as a valid diagnostic of PAH size, and the effectiveness of this diagnostic is  governed by aliphatic content.
Each fixed-$N_{\rm C}$ sequence (connecting neutral and cationic points) is well described by a parabola, yet the curvature of these parabolic fits varies dramatically with $N_{\rm C}$.
Figure~\ref{fig:fig6} is  obtained from Figure~\ref{fig:fig4} by taking vertical tracks (which appear as near-vertical or parabolic loci in the ratio plane) at fixed carbon atom numbers $N_{\mathrm{c}}$ (i.e., fixed PAH size bins). For a given $N_{\mathrm{c}}$, we examine how the grid position shifts with increasing aliphatic fraction, thereby isolating the aliphatic effect at a given molecular size.
Small PAHs (e.g., $N_{\rm C}=26$ or 48) exhibit  strongly bent curves in the $I_{11.2}/I_{7.7}$ versus $I_{11.2}/I_{3.3}$ plane, and these curves shift substantially as the aliphatic fraction increases. In contrast, large PAHs ($N_{\rm C}=76$ or 98) produce nearly monotonic, overlapping curves; their neutral–cation sequences are almost linear and show minimal change with aliphatic fractions. Quantitatively, the curvature coefficient ($A$ in Equation~\ref{avalue}) of the fitted parabola is highest for the smallest molecules (peaking at $\sim$10 for a fully aliphatic $N_{\rm C}=26$ PAH) and drops by an order of magnitude towards the largest sizes (down to $A \lesssim 1$ for $N_{\rm C}=98$; see Figure~\ref{fig:fig6} bar chart). Thus, the inclusion of aliphatic structure induces a strong ``diagnostic distortion'' for small PAHs, while the diagnostic ratios of large PAHs remain relatively insensitive to aliphatic modifications.
A direct comparison more clearly illustrates this size dependence. For the small-size track at $N_{\mathrm C}\approx 26$, the N50C50 point shifts from about $\log(I_{11.2}/I_{3.3})\approx -0.37$ and $\log(I_{11.2}/I_{7.7})\approx -0.44$ in the purely aromatic case to $\log(I_{11.2}/I_{3.3})\approx -0.53$ and $\log(I_{11.2}/I_{7.7})\approx -0.71$ in the 100\% aliphatic case. In contrast, for the large-size track at $N_{\mathrm C}\approx 97$, the corresponding N50C50 point shows only a weak variation, from roughly $\log(I_{11.2}/I_{3.3})\approx 0.87$ and $\log(I_{11.2}/I_{7.7})\approx -0.01$ to $\log(I_{11.2}/I_{3.3})\approx 0.85$ and $\log(I_{11.2}/I_{7.7})\approx -0.01$. In short, the diagnostic distortion introduced by aliphatic structure is substantial for small PAHs but becomes much less significant for large PAHs.

These trends can be explained by examining the vibrational properties of small versus large PAHs and how aliphatic substitution modifies their IR emission. Small PAHs possess a limited number of vibration modes and are primarily governed by edge-linked C–H oscillators \citep{1989ApJS...71..733A}. 
The most prominent C–H vibrational mode of small PAHs corresponds to the C–H stretching modes in the 3 $\mu$m region.
The out-of-plane emission in the 11--15 $\mu$m 
 range is generally distributed across duo, trio, and quarto C–H bending modes, rather than being dominated by the 11.2 $\mu$m solo component. This is because small PAHs typically lack long straight edges and thus possess relatively few solo sites
\citep{2004ASPC..309..665H}. Consequently, the intensity ratio $I_{11.2}/I_{3.3}$ decreases with decreasing $N_{\rm C}$. 
 Even minor structural modifications, such as converting a few peripheral H atoms to aliphatic bonds, substantially redistribute IR intensity \citep{2013ApJS..205....8S}. Laboratory spectra confirm that partial hydrogenation of PAHs induces two major shifts in the spectrum:  (1) the aromatic 3.3 $\mu$m C–H stretch weakens, while a new aliphatic stretch around 3.4 $\mu$m grows, and (2) 
the aromatic 11--15 $\mu$m C–H oop bands shift and weaken alongside the emergence of a strong aliphatic C–H deformation band at 6.9 $\mu$m \citep{2013ApJS..205....8S}. 
In other words, introducing aliphatic side groups or superhydrogenating a small PAH tends to suppress  emission at 3.3 $\mu$m and 11.2 $\mu$m,  while enhancing new aliphatic vibrational modes.
This induces a systematic shift in the  $I_{11.2}/I_{7.7}$ versus $I_{11.2}/I_{3.3}$ plot, as shown in Figure~\ref{fig:fig6}.

Vibrational mode coupling probably amplifies this effect. In small molecules, the aromatic C–H stretching and bending modes are relatively isolated; thus, introducing an aliphatic –CH$_2$- or –CH$_3$ group can effectively transfer intensity from these aromatic modes into newly accessible coupling pathways and overtone or combination transitions.
In fact, theoretical calculations show that the combination bands, especially the 5--6 $\mu$m plateau features, carry only a modest fraction of the vibrational energy in small PAHs but become increasingly important as the PAH size increases \citep{2019A&A...628A.130L}. Thus, a small PAH  channels a larger portion of its energy into the 3.3 $\mu$m and 11.2 $\mu$m fundamentals; any structural change that affects those modes yields a large shift in the observed ratios.

For large PAHs, the total number of solo C–H bonds increases; consequently, the incorporation of aliphatic components may exert comparable effects on the 3.3 $\mu$m and 11.2 $\mu$m emission bands. Moreover, introducing an aliphatic group into a large PAH modifies only a small fraction of its total C–H bonds. As a result, any incremental variation in the intensity of the 3.3 $\mu$m or 11.2 $\mu$m bands is diluted by the extensive aromatic scaffold of the large molecule. It follows that the $I_{11.2}/I_{3.3}$ intensity ratio for large PAHs is scarcely altered by the introduction of aliphatic components. For ionized PAHs, the strength of the C–C vibrational mode is markedly enhanced relative to that of the C–H mode, and the overall spectral profile becomes less sensitive to peripheral (aliphatic) modifications, even for small PAHs (corresponding to the low $I_{11.2}/I_{7.7}$ range, as illustrated in Figure~\ref{fig:fig6}).

In summary, diagnostic grid analysis demonstrates that the PAH size is the dominant factor governing the shape of charge–intensity diagrams, with the aliphatic fraction acting as a secondary modifier that is most impactful at small sizes. Small PAHs experience significant spectral distortion in diagnostic ratios when aliphatic side groups are present, making their mid-IR band ratios difficult to interpret solely through size or charge considerations. In contrast, large PAHs maintain relatively stable diagnostic ratios despite a substantial aliphatic content. From an observational standpoint, caution is warranted when applying PAH diagnostic ratios in environments where very small or irregular PAHs prevail, such as in circumstellar environments of cool stars or transient hydrocarbon clusters, as their spectral behavior may deviate from the calibrated grids derived for ideal PAHs. However, in regimes dominated by large PAHs, such as diffuse ISM or mature PDRs, conventional diagnostic diagrams remain reliable, as these PAHs inherently exhibit greater spectral resilience.

 \begin{figure*}
    \includegraphics[width=\textwidth]{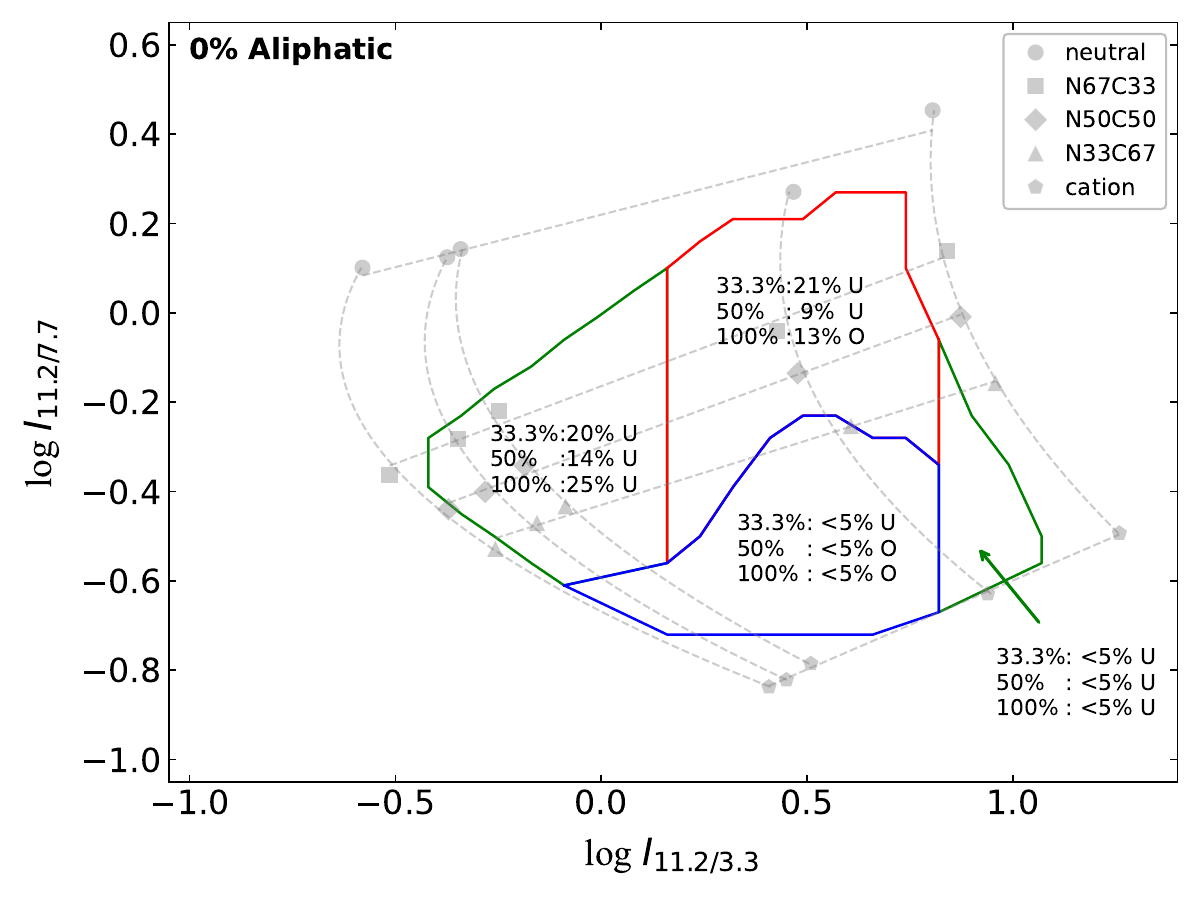}
    \caption{Bias map of the inferred ionization fraction when PAH populations containing aliphatic components are interpreted using the 0\% aliphatic (purely aromatic) diagnostic grid. The gray dashed curves and symbols show the reference grid for the 0\% aliphatic case. The colored regions mark zones with different bias characteristics. The text labels inside each region summarize the deviation for the 33.3\%, 50\%, and 100\% aliphatic cases relative to the purely aromatic interpretation.
    Here, the ionization fraction is parameterized from 0 for purely neutral populations to 1 for purely cationic populations; thus, ``U'' and ``O'' denote underestimation and overestimation of the inferred ionization fraction, and ``$<5\%$'' indicates that the deviation is smaller than 5\%. For example, the label ``33.3\%: 20\% U'' means that, in that region, the ionization fraction inferred for the 33.3\% aliphatic case is underestimated by about 20\% if the source is interpreted using the aromatic-only grid.}
    \label{fig:fig7}   
\end{figure*}

 \begin{figure*}
    \includegraphics[width=\textwidth]{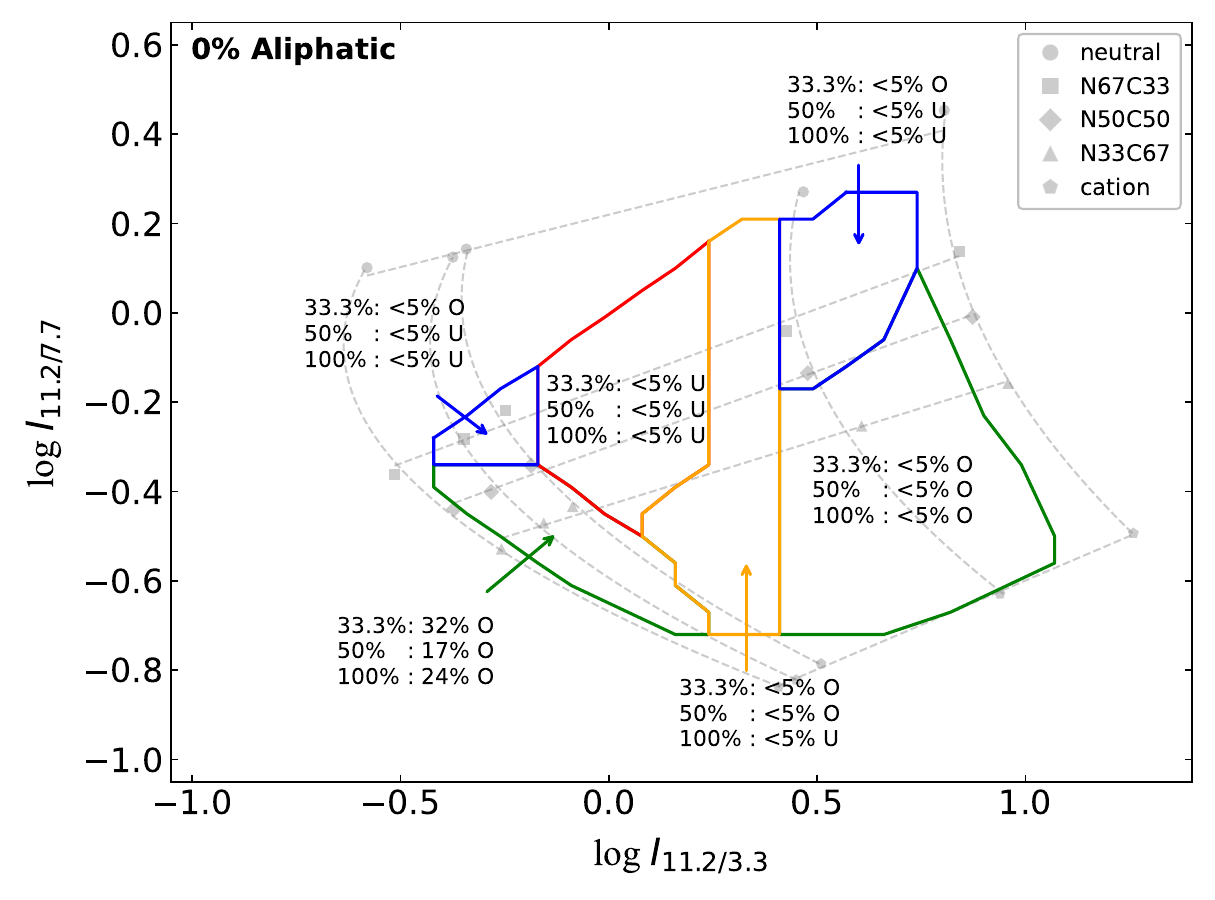}
    \caption{Bias map of the inferred characteristic PAH size, expressed as carbon atom number $N_{\mathrm{C}}$, when PAH populations containing aliphatic components are interpreted using the 0\% aliphatic (purely aromatic) diagnostic grid. The gray dashed curves and symbols show the reference grid for the 0\% aliphatic case. The colored regions mark zones with different bias characteristics. The text labels inside each region summarize the deviation for the 33.3\%, 50\%, and 100\% aliphatic cases relative to the purely aromatic interpretation. Here, ``U'' denotes an underestimated carbon atom number, ``O'' denotes an overestimated carbon atom number, and ``$<5\%$'' indicates that the deviation is smaller than 5\%. For example, the label ``33.3\%: 32\% O'' means that, in that region, the characteristic PAH size inferred for the 33.3\% aliphatic case is overestimated by about 32\% if the source is interpreted using the aromatic-only grid.}
    \label{fig:fig8}
\end{figure*}

\subsection{Quantifying the effects of aliphatic components}

By applying linear interpolation across the overlapping domains of the four diagnostic grids (0$\%$, 33.3$\%$, 50$\%$, and 100$\%$ aliphatic carbon fractions), we quantified the deviations of each population rich in aliphatic from the purely aromatic baseline at every point in the band-ratio space.
For the ionization-fraction bias map, we assign numerical values of 0 and 1 to purely neutral and purely cationic populations, respectively, and estimate intermediate ionization fractions by interpolation within the diagnostic grid.

The resulting difference maps,showing ionization fraction in Figure \ref{fig:fig7} and $N_{\rm C}$ in Figure \ref{fig:fig8}, are partitioned into color-coded regions that represent varying biases between the purely aromatic PAHs and those containing different aliphatic fractions. For the ionization fraction (Figure \ref{fig:fig7}), the green-shaded regions indicate a consistent underestimation of ionization in the three aliphatic fractions relative to purely aromatic PAHs, while in other regions, the biases vary, with some aliphatic fractions resulting in overestimation and others in underestimation. Similarly, for $N_{\rm C}$ (Figure \ref{fig:fig8}), the green-shaded areas correspond to a consistent overestimation of PAH size for all aliphatic fractions compared to aromatic-only PAHs, while other regions exhibit mixed biases depending on the specific aliphatic fraction. The quantitative deviations are indicated directly in the figures.
Notably, the largest deviations occur within the leftmost, green-shaded regions of both figures, where all aliphatic-bearing models lead to systematic underestimations of the PAH ionization fraction and overestimations of PAH size. In this zone, the inferred ionization fraction is underestimated by about 20$\%$, $14\%$ and $25\%$ for the 33.3$\%$, 50$\%$ and 100$\%$ aliphatic fractions, respectively. Similarly, the inferred PAH size  is overestimated by approximately 32$\%$, 17$\%$, and 24$\%$ for the 33.3$\%$, 50$\%$, and 100$\%$ aliphatic cases, respectively. These significant biases highlight that even a moderate aliphatic content can substantially distort PAH diagnostics when an entirely aromatic spectral template is assumed.

Observationally, this effect is  linked to the strength of the UV radiation field.  In weaker-UV environments (such as around B-type stars or reddened C-type sources), PAHs tend to retain a large aliphatic fraction, producing stronger 11.2 $\mu$m and 6.9 and 7.25 $\mu$m emission bands. In contrast, in strong UV environments (e.g. around A-type stars), PAHs become almost fully aromatized \citep{2013ApJS..205....8S}. This UV-dependent aromatization pathway can be probed by analyzing the observational data associated with the spatial distribution of the grids: the left (green) regions correspond to low-UV conditions dominated by aliphatic-rich PAHs, while the right regions represent high-UV environments favoring purely aromatic PAHs.

Recognizing molecular structure as a third parameter (alongside size and charge) will enhance the accuracy of PAH-based astrophysical diagnostics. At minimum, future frameworks should identify regimes (such as the ``left green zone'' in Figure \ref{fig:fig7} and Figure \ref{fig:fig8}) where standard aromatic-based calibrations may be inaccurate.

\subsection{Applications in observational data analysis}

Based on the observational data of NGC 7023 reported in \cite{2015A&A...577A..16P} and \cite{MPR2020}, we estimated the properties of PAHs  by inputting these  data into our diagnostic grid. Our analysis reveals that incorporating a fraction of aliphatic carbon 
induces a significant deviation in the inferred results, compared to those derived from
a purely aromatic grid. When applying an aromatic-only model, 
the mid-infrared emission features of NGC 7023 indicate  
a high ionization fraction ($\sim$ 0.8, meaning approximately 80$\%$ of PAHs exist as cations) and an average molecular size of  $N_{\rm C} \approx 75 \pm 10$ carbon atoms. 
In contrast, accounting for a  moderate aliphatic fraction yields a slightly lower ionization fraction ($\sim$0.73) and a larger derived molecular size ($N_{\rm C} \approx 80 \pm 10$). 
Consequently, the presence of aliphatics can create the misleading impression that the PAH population is more neutral and larger than it actually is.
Observational evidence from NGC 7023 further supports this effect. First, the 3.4 $\mu$m aliphatic C–H feature is notably stronger in the shielded core than at its UV-exposed interface; accordingly, the $I_{3.3}/I_{3.4}$ $\mu$m intensity ratio decreases from $\sim$0.13 to 0.03 in this transition \citep{2015A&A...577A..16P}. Second, the  $I_{11.2}/I_{3.3}$ $\mu$m 
intensity ratio reaches its minimum at the PDR interface of NGC 7023, where small, highly ionized PAHs dominate. As one moves deeper into the nebula, this ratio increases, accompanying
a shift in the PAH population toward larger, more neutral molecules
 \citep{2016A&A...590A..26C}. 
If aliphatic components were omitted from the analysis, the enhanced 11.2 $\mu$m emission
in the nebular interior would be incorrectly attributed solely to a higher abundance of
large neutral PAHs, ultimately skewing the inferred PAH charge state and size distribution.

While NGC 7023 exemplifies how a moderate aliphatic content can substantially alter the inferred size and charge of PAHs, the case of M82 offers a clear contrast.
In the starburst galaxy M82, where PAH emission is integrated across numerous star-forming regions, the purely aromatic diagnostic grid already positions its PAH population near the extreme cationic end of the charge–size diagram \citep{MPR2020}. 
Our purely aromatic diagnostic grid yields an inferred average PAH size of approximately
$N_{\rm C}\approx 88 \pm 13$ carbon atoms. Presumably,  
smaller PAHs have been destroyed by the galactic superwind, leaving only large, resilient molecules intact. When aliphatic content is incorporated into the models, the derived ionization fraction and average PAH size undergo only minor adjustment, typically just 
a few percent.

\section{Conclusions}\label{sec:5}
In this work, we investigated the influence of aliphatic side groups on the IR emission features of PAHs. We analyzed a broad set of PAH molecules (both purely aromatic and those containing aliphatic substituents) using simulated IR spectra. This approach allowed us to quantify how the addition of aliphatic components alters key diagnostic band intensity ratios that are commonly used to infer the PAH size and charge.

Our results show that aliphatic functional groups systematically alter PAH band intensity ratios, notably by modifying $I_{11.2}/I_{7.7}$ (an indicator of PAH charge state) and $I_{11.2}/I_{3.3}$ (an indicator of  PAH size). 
 Failure to account for the effects of aliphatic components could thus lead to misclassification of PAH charge states. The magnitude of distortion caused by aliphatic side groups is strongly size-dependent: small PAHs exhibit much larger perturbations in their IR band ratios when aliphatic groups are attached, whereas large PAHs are only mildly affected. 
 
 Quantitatively, the neglect of aliphatic components in PAH diagnostics leads to systematic biases in the inferred physical properties, with the most significant deviations occurring in the left region of the 
$I_{11.2}/I_{7.7}$ versus $I_{11.2}/I_{3.3}$ diagnostic grid, corresponding to low-UV, aliphatic-rich astrophysical environments (e.g., shielded cores of photodissociation regions, circumstellar envelopes of cool stars).
For these regions, adopting a purely aromatic calibration underestimates the PAH ionization fraction by 14-–25\% and overestimates the molecular size  by 17--32\% for aliphatic fractions ranging from 33.3\% to 100\%. 
 In other regions of the diagnostic grid, the biases are typically less than 5\%, indicating that the impact of aliphatic components is spatially and environmentally dependent in observational contexts. 
 Despite these structural perturbations, the conventional diagnostic grid in the $I_{11.2}/I_{7.7}$ versus $I_{11.2}/I_{3.3}$ plane remains largely robust. The general framework of this diagnostic diagram is still applicable for interpreting PAH populations.
This preservation of relative trends means the traditional diagnostic framework is still applicable for aliphatic-containing PAHs, provided that aliphatic fraction is introduced as a critical third parameter (alongside size and ionization state) for calibration and correction.
 
Our results also highlight the observational implications of aliphatic functionalization for interpreting PAH emission spectra, particularly for high-resolution JWST data. For the well-studied PDR NGC 7023, accounting for a moderate aliphatic fraction revises the inferred PAH ionization fraction from $\sim0.8$ (purely aromatic model) to $\sim0.73$ and the average molecular size from $N_{\rm C}=75\pm10$ to $80\pm10$, demonstrating that aliphatic components can create the misleading impression of a more neutral and larger PAH population if unaccounted for. In contrast, for the starburst galaxy M82, an environment dominated by strong UV radiation and fully aromatized PAHs, the inclusion of aliphatic content induces only minor ($<5\%$) adjustments to the inferred PAH properties. These case studies illustrate the necessity of tailoring PAH diagnostic analyses to the specific astrophysical environment, with aliphatic correction being essential for low-UV, shielded regions and less critical for extreme UV-exposed environments.

Beyond the core findings, this work also identifies key methodological limitations that provide important context for interpreting our results and guiding future research. First, our PAH sample is constrained by gaps in the  PAHdb. Neutral aliphatic PAHs are absent in the $N_{\rm C}$ ranges of 30--45 and 61--85, and cationic aliphatic PAHs show missing data in 35--50 and 61--95 $N_{\rm C}$, which may leave some size-dependent aliphatic effects unaccounted for. Second, we employed a simplified radiation field model with an average photon energy of 6 eV to generate PAH emission spectra, which does not capture the full complexity of interstellar radiation fields (e.g., spatial gradients in the interstellar radiation field, local UV enhancement from massive stars). Third, we did not subdivide aliphatic-bearing PAHs by their structural modification type (superhydrogenation vs. methyl/methylene substitution) due to insufficient sample size per subtype, precluding an assessment of the differentiated impact of distinct aliphatic functional groups on PAH emission ratios. Finally, while we conducted anharmonicity-related tests and confirmed that anharmonic effects primarily rescale the absolute size calibration (rather than erasing aromatic-versus-aliphatic trends), our analysis is based on harmonic emission spectra, and a fully anharmonic treatment of aliphatic-bearing PAHs remains unimplemented.

Despite these limitations, our study establishes a preliminary framework for PAH diagnostic analyses accounting for aliphatic structural diversity and identifies key future research directions, such as optimizing the aliphatic PAH database, developing complex radiation field models and building a 3D PAH diagnostic framework integrating size, ionization state and aliphatic fraction. In summary, this work verifies that aliphatic components are essential for interpreting PAH emission band ratios—their neglect causes notable inference biases in aliphatic-rich environments. By introducing aliphatic fraction as a third diagnostic parameter, we move beyond the purely aromatic PAH paradigm, providing critical calibration for JWST high-resolution PAH spectral analysis, facilitating accurate characterization of interstellar PAH populations and deeper insights into interstellar organic molecular evolution, while highlighting the need for synergy between theoretical modeling, laboratory spectroscopy and high-resolution astronomical observations.

\section*{Acknowledgements}

We thank two anonymous referees for insightful comments that have greatly enhanced this paper.
The financial supports of this work are from 
the National SKA Program of China (No.\,2025SKA0120100), 
the National Natural Science Foundation of China (NSFC, No.\,12473027 and 12333005), the Guangdong Basic and Applied Basic Research Funding (No.\,2024A1515010798), and the Greater Bay Area Branch of the National Astronomical Data Center (No.\,2024B1212080003). 
This article is based
upon work from COST Action CA21126 - Carbon molecular nanostructures in space (NanoSpace), supported by COST (European Cooperation in Science
and Technology).

\section*{Data Availability} 
The analysis products of this work will be shared on a reasonable
request to the corresponding author.



\bibliographystyle{mnras}
\bibliography{reference} 




\appendix

\section{PAH uid list}\label{uid list}
Table \ref{tab:tabA1} and \ref{tab:tabA2} lists the UIDs of aliphatic-containing PAH molecules, including both neutral and cationic species, selected from version 3.20 of the NASA Ames PAH IR Spectroscopic Database (PAHdb). Table \ref{tab:tabA3} and \ref{tab:tabA4} provides the corresponding UIDs for purely aromatic PAH molecules, also including both neutral and cationic forms, from the same database version.
 \begin{table*}
	\centering
	\caption{PAHdb v3.20 UIDs for neutral PAHs with aliphatic chains.}
	\label{tab:tabA1}
	\begin{tabular}{lcccccccccccccccccc} 
		\hline
		PAH  & UID & & & PAH  & UID & & & PAH  & UID & & & PAH  & UID & & & PAH  & UID \\
		\hline
$\mathrm {C_{	22	}H_{	16	}}$	&	356	&&&	$\mathrm {C_{	27	}H_{	14	}}$	&	72	&&&	$\mathrm {C_{	54	}H_{	20	}}$	&	823	&&&	$\mathrm {C_{	59	}H_{	20	}}$	&	93	&&&	$\mathrm {C_{	96	}H_{	28	}}$	&	189	\\
$\mathrm {C_{	24	}H_{	16	}}$	&	378	&&&	$\mathrm {C_{	27	}H_{	14	}}$	&	73	&&&	$\mathrm {C_{	54	}H_{	18	}}$	&	828	&&&	$\mathrm {C_{	59	}H_{	20	}}$	&	94	&&&	$\mathrm {C_{	96	}H_{	28	}}$	&	190	\\
$\mathrm {C_{	24	}H_{	16	}}$	&	379	&&&	$\mathrm {C_{	47	}H_{	18	}}$	&	78	&&&	$\mathrm {C_{	55	}H_{	20	}}$	&	316	&&&	$\mathrm {C_{	59	}H_{	20	}}$	&	95	&&&	$\mathrm {C_{	96	}H_{	25	}}$	&	682	\\
$\mathrm {C_{	24	}H_{	16	}}$	&	380	&&&	$\mathrm {C_{	47	}H_{	18	}}$	&	79	&&&	$\mathrm {C_{	55	}H_{	20	}}$	&	318	&&&	$\mathrm {C_{	59	}H_{	20	}}$	&	96	&&&	$\mathrm {C_{	96	}H_{	25	}}$	&	683	\\
$\mathrm {C_{	24	}H_{	16	}}$	&	381	&&&	$\mathrm {C_{	47	}H_{	18	}}$	&	80	&&&	$\mathrm {C_{	56	}H_{	20	}}$	&	806	&&&	$\mathrm {C_{	59	}H_{	20	}}$	&	97	&&&	$\mathrm {C_{	97	}H_{	26	}}$	&	799	\\
$\mathrm {C_{	25	}H_{	14	}}$	&	314	&&&	$\mathrm {C_{	47	}H_{	18	}}$	&	81	&&&	$\mathrm {C_{	56	}H_{	20	}}$	&	807	&&&	$\mathrm {C_{	59	}H_{	20	}}$	&	98	&&&	$\mathrm {C_{	97	}H_{	26	}}$	&	800	\\
$\mathrm {C_{	27	}H_{	14	}}$	&	67	&&&	$\mathrm {C_{	47	}H_{	18	}}$	&	82	&&&	$\mathrm {C_{	57	}H_{	22	}}$	&	808	&&&	$\mathrm {C_{	59	}H_{	20	}}$	&	99	&&&	$\mathrm {C_{	98	}H_{	26	}}$	&	801	\\
$\mathrm {C_{	27	}H_{	14	}}$	&	68	&&&	$\mathrm {C_{	47	}H_{	18	}}$	&	83	&&&	$\mathrm {C_{	58	}H_{	22	}}$	&	809	&&&	$\mathrm {C_{	59	}H_{	24	}}$	&	810	&&&	$\mathrm {C_{	99	}H_{	28	}}$	&	802	\\
$\mathrm {C_{	27	}H_{	14	}}$	&	69	&&&	$\mathrm {C_{	47	}H_{	18	}}$	&	84	&&&	$\mathrm {C_{	59	}H_{	20	}}$	&	90	&&&	$\mathrm {C_{	60	}H_{	24	}}$	&	811	&&&	$\mathrm {C_{	100	}H_{	28	}}$	&	803	\\
$\mathrm {C_{	27	}H_{	14	}}$	&	70	&&&	$\mathrm {C_{	47	}H_{	18	}}$	&	85	&&&	$\mathrm {C_{	59	}H_{	20	}}$	&	91	&&&	$\mathrm {C_{	60	}H_{	24	}}$	&	812	&&&	$\mathrm {C_{	101	}H_{	30	}}$	&	804	\\
$\mathrm {C_{	27	}H_{	14	}}$	&	71	&&&	$\mathrm {C_{	47	}H_{	18	}}$	&	86	&&&	$\mathrm {C_{	59	}H_{	20	}}$	&	92	&&&	$\mathrm {C_{	87	}H_{	24	}}$	&	650	&&&	$\mathrm {C_{	102	}H_{	30	}}$	&	805	\\

        \hline
	\end{tabular}
\end{table*}

 \begin{table*}
	\centering
	\caption{PAHdb v3.20 UIDs for cationic PAHs with aliphatic chains.}
	\label{tab:tabA2}
	\begin{tabular}{lcccccccccccccccccc} 
		\hline
		PAH  & UID & & & PAH  & UID & & & PAH  & UID & & & PAH  & UID & & & PAH  & UID \\
		\hline
$\mathrm {C_{	22	}H_{	16	}^{+}}$	&	371	&&&	$\mathrm {C_{	54	}H_{	21	}^{+}}$	&	213	&&&	$\mathrm {C_{	56	}H_{	20	}^{+}}$	&	791	&&&	$\mathrm {C_{	96	}H_{	28	}^{+}}$	&	191	&&&	$\mathrm {C_{	99	}H_{	28	}^{+}}$	&	780	\\
$\mathrm {C_{	24	}H_{	13	}^{+}}$	&	34	&&&	$\mathrm {C_{	54	}H_{	21	}^{+}}$	&	214	&&&	$\mathrm {C_{	57	}H_{	22	}^{+}}$	&	792	&&&	$\mathrm {C_{	96	}H_{	47	}^{+}}$	&	587	&&&	$\mathrm {C_{	99	}H_{	27	}^{+}}$	&	781	\\
$\mathrm {C_{	25	}H_{	14	}^{+}}$	&	315	&&&	$\mathrm {C_{	54	}H_{	35	}^{+}}$	&	585	&&&	$\mathrm {C_{	57	}H_{	21	}^{+}}$	&	793	&&&	$\mathrm {C_{	96	}H_{	47	}^{+}}$	&	588	&&&	$\mathrm {C_{	100	}H_{	28	}^{+}}$	&	782	\\
$\mathrm {C_{	32	}H_{	15	}^{+}}$	&	6	&&&	$\mathrm {C_{	54	}H_{	35	}^{+}}$	&	586	&&&	$\mathrm {C_{	58	}H_{	22	}^{+}}$	&	794	&&&	$\mathrm {C_{	96	}H_{	29	}^{+}}$	&	589	&&&	$\mathrm {C_{	101	}H_{	30	}^{+}}$	&	783	\\
$\mathrm {C_{	32	}H_{	15	}^{+}}$	&	7	&&&	$\mathrm {C_{	54	}H_{	18	}^{+}}$	&	827	&&&	$\mathrm {C_{	59	}H_{	24	}^{+}}$	&	795	&&&	$\mathrm {C_{	96	}H_{	33	}^{+}}$	&	590	&&&	$\mathrm {C_{	101	}H_{	29	}^{+}}$	&	784	\\
$\mathrm {C_{	32	}H_{	15	}^{+}}$	&	8	&&&	$\mathrm {C_{	55	}H_{	20	}^{+}}$	&	317	&&&	$\mathrm {C_{	59	}H_{	23	}^{+}}$	&	796	&&&	$\mathrm {C_{	97	}H_{	26	}^{+}}$	&	775	&&&	$\mathrm {C_{	102	}H_{	30	}^{+}}$	&	785	\\
$\mathrm {C_{	32	}H_{	15	}^{+}}$	&	9	&&&	$\mathrm {C_{	55	}H_{	20	}^{+}}$	&	319	&&&	$\mathrm {C_{	60	}H_{	24	}^{+}}$	&	797	&&&	$\mathrm {C_{	97	}H_{	26	}^{+}}$	&	776	&&&	$\mathrm {C_{	108	}H_{	36	}^{+}}$	&	786	\\
$\mathrm {C_{	54	}H_{	19	}^{+}}$	&	40	&&&	$\mathrm {C_{	55	}H_{	19	}^{+}}$	&	788	&&&	$\mathrm {C_{	60	}H_{	24	}^{+}}$	&	798	&&&	$\mathrm {C_{	97	}H_{	25	}^{+}}$	&	777	&&&	$\mathrm {C_{	108	}H_{	36	}^{+}}$	&	787	\\
$\mathrm {C_{	54	}H_{	19	}^{+}}$	&	41	&&&	$\mathrm {C_{	55	}H_{	19	}^{+}}$	&	789	&&&	$\mathrm {C_{	96	}H_{	25	}^{+}}$	&	109	&&&	$\mathrm {C_{	97	}H_{	25	}^{+}}$	&	778	&&&	$\mathrm {C_{	112	}H_{	27	}^{+}}$	&	609	\\
$\mathrm {C_{	54	}H_{	21	}^{+}}$	&	212	&&&	$\mathrm {C_{	56	}H_{	20	}^{+}}$	&	790	&&&	$\mathrm {C_{	96	}H_{	25	}^{+}}$	&	110	&&&	$\mathrm {C_{	98	}H_{	26	}^{+}}$	&	779	&&&	$\mathrm {C_{	112	}H_{	27	}^{+}}$	&	610	\\

		\hline
	\end{tabular}
\end{table*}

 \begin{table*}
	\centering
	\caption{PAHdb v3.20 UIDs for neutral PAHs without aliphatic chains.}
	\label{tab:tabA3}
	\begin{tabular}{lcccccccccccccccccc} 
		\hline
		PAH  & UID & & & PAH  & UID & & & PAH  & UID & & & PAH  & UID & & & PAH  & UID \\
		\hline
$\mathrm {C_{	22	}H_{	14	}}$	&	301	&&&	$\mathrm {C_{	27	}H_{	15	}}$	&	2245	&&&	$\mathrm {C_{	52	}H_{	17	}}$	&	835	&&&	$\mathrm {C_{	59	}H_{	21	}}$	&	3247	&&&	$\mathrm {C_{	88	}H_{	24	}}$	&	3206	\\
$\mathrm {C_{	22	}H_{	14	}}$	&	305	&&&	$\mathrm {C_{	27	}H_{	15	}}$	&	2247	&&&	$\mathrm {C_{	52	}H_{	18	}}$	&	3173	&&&	$\mathrm {C_{	60	}H_{	28	}}$	&	3647	&&&	$\mathrm {C_{	90	}H_{	30	}}$	&	171	\\
$\mathrm {C_{	22	}H_{	14	}}$	&	2340	&&&	$\mathrm {C_{	27	}H_{	15	}}$	&	2248	&&&	$\mathrm {C_{	53	}H_{	16	}}$	&	826	&&&	$\mathrm {C_{	60	}H_{	28	}}$	&	3658	&&&	$\mathrm {C_{	90	}H_{	24	}}$	&	618	\\
$\mathrm {C_{	22	}H_{	14	}}$	&	2342	&&&	$\mathrm {C_{	27	}H_{	15	}}$	&	2249	&&&	$\mathrm {C_{	54	}H_{	18	}}$	&	37	&&&	$\mathrm {C_{	60	}H_{	26	}}$	&	3706	&&&	$\mathrm {C_{	90	}H_{	24	}}$	&	638	\\
$\mathrm {C_{	22	}H_{	14	}}$	&	2343	&&&	$\mathrm {C_{	46	}H_{	18	}}$	&	3169	&&&	$\mathrm {C_{	54	}H_{	18	}}$	&	836	&&&	$\mathrm {C_{	60	}H_{	26	}}$	&	3721	&&&	$\mathrm {C_{	90	}H_{	28	}}$	&	3368	\\
$\mathrm {C_{	22	}H_{	14	}}$	&	2344	&&&	$\mathrm {C_{	47	}H_{	17	}}$	&	76	&&&	$\mathrm {C_{	54	}H_{	20	}}$	&	3176	&&&	$\mathrm {C_{	60	}H_{	26	}}$	&	3722	&&&	$\mathrm {C_{	96	}H_{	25	}}$	&	684	\\
$\mathrm {C_{	22	}H_{	14	}}$	&	2345	&&&	$\mathrm {C_{	48	}H_{	18	}}$	&	35	&&&	$\mathrm {C_{	54	}H_{	26	}}$	&	3850	&&&	$\mathrm {C_{	60	}H_{	26	}}$	&	3725	&&&	$\mathrm {C_{	96	}H_{	25	}}$	&	685	\\
$\mathrm {C_{	24	}H_{	12	}}$	&	18	&&&	$\mathrm {C_{	48	}H_{	20	}}$	&	100	&&&	$\mathrm {C_{	54	}H_{	26	}}$	&	3856	&&&	$\mathrm {C_{	60	}H_{	26	}}$	&	3726	&&&	$\mathrm {C_{	96	}H_{	25	}}$	&	686	\\
$\mathrm {C_{	24	}H_{	14	}}$	&	204	&&&	$\mathrm {C_{	48	}H_{	22	}}$	&	143	&&&	$\mathrm {C_{	54	}H_{	24	}}$	&	3868	&&&	$\mathrm {C_{	60	}H_{	26	}}$	&	3728	&&&	$\mathrm {C_{	96	}H_{	25	}}$	&	687	\\
$\mathrm {C_{	24	}H_{	14	}}$	&	208	&&&	$\mathrm {C_{	48	}H_{	20	}}$	&	146	&&&	$\mathrm {C_{	54	}H_{	24	}}$	&	3869	&&&	$\mathrm {C_{	60	}H_{	26	}}$	&	3730	&&&	$\mathrm {C_{	96	}H_{	25	}}$	&	688	\\
$\mathrm {C_{	24	}H_{	14	}}$	&	2330	&&&	$\mathrm {C_{	48	}H_{	18	}}$	&	635	&&&	$\mathrm {C_{	54	}H_{	24	}}$	&	3870	&&&	$\mathrm {C_{	60	}H_{	26	}}$	&	3731	&&&	$\mathrm {C_{	96	}H_{	25	}}$	&	690	\\
$\mathrm {C_{	24	}H_{	14	}}$	&	2331	&&&	$\mathrm {C_{	48	}H_{	24	}}$	&	3931	&&&	$\mathrm {C_{	54	}H_{	24	}}$	&	3871	&&&	$\mathrm {C_{	60	}H_{	26	}}$	&	3732	&&&	$\mathrm {C_{	96	}H_{	26	}}$	&	691	\\
$\mathrm {C_{	24	}H_{	14	}}$	&	2332	&&&	$\mathrm {C_{	48	}H_{	24	}}$	&	3932	&&&	$\mathrm {C_{	54	}H_{	24	}}$	&	3872	&&&	$\mathrm {C_{	60	}H_{	26	}}$	&	3734	&&&	$\mathrm {C_{	96	}H_{	26	}}$	&	692	\\
$\mathrm {C_{	24	}H_{	14	}}$	&	2333	&&&	$\mathrm {C_{	48	}H_{	22	}}$	&	3937	&&&	$\mathrm {C_{	54	}H_{	24	}}$	&	3876	&&&	$\mathrm {C_{	60	}H_{	26	}}$	&	3739	&&&	$\mathrm {C_{	96	}H_{	23	}}$	&	693	\\
$\mathrm {C_{	24	}H_{	14	}}$	&	2334	&&&	$\mathrm {C_{	48	}H_{	22	}}$	&	3939	&&&	$\mathrm {C_{	54	}H_{	24	}}$	&	3885	&&&	$\mathrm {C_{	60	}H_{	26	}}$	&	3740	&&&	$\mathrm {C_{	96	}H_{	23	}}$	&	696	\\
$\mathrm {C_{	24	}H_{	14	}}$	&	2335	&&&	$\mathrm {C_{	48	}H_{	22	}}$	&	3940	&&&	$\mathrm {C_{	54	}H_{	22	}}$	&	3924	&&&	$\mathrm {C_{	60	}H_{	24	}}$	&	3792	&&&	$\mathrm {C_{	96	}H_{	22	}}$	&	699	\\
$\mathrm {C_{	25	}H_{	13	}}$	&	2323	&&&	$\mathrm {C_{	48	}H_{	22	}}$	&	3941	&&&	$\mathrm {C_{	54	}H_{	22	}}$	&	3925	&&&	$\mathrm {C_{	60	}H_{	24	}}$	&	3793	&&&	$\mathrm {C_{	98	}H_{	28	}}$	&	567	\\
$\mathrm {C_{	26	}H_{	14	}}$	&	2291	&&&	$\mathrm {C_{	48	}H_{	22	}}$	&	3942	&&&	$\mathrm {C_{	55	}H_{	19	}}$	&	3244	&&&	$\mathrm {C_{	60	}H_{	24	}}$	&	3794	&&&	$\mathrm {C_{	99	}H_{	25	}}$	&	3282	\\
$\mathrm {C_{	26	}H_{	14	}}$	&	2300	&&&	$\mathrm {C_{	48	}H_{	22	}}$	&	3943	&&&	$\mathrm {C_{	56	}H_{	20	}}$	&	3179	&&&	$\mathrm {C_{	60	}H_{	24	}}$	&	3795	&&&	$\mathrm {C_{	102	}H_{	26	}}$	&	177	\\
$\mathrm {C_{	27	}H_{	13	}}$	&	66	&&&	$\mathrm {C_{	48	}H_{	22	}}$	&	3944	&&&	$\mathrm {C_{	57	}H_{	19	}}$	&	646	&&&	$\mathrm {C_{	60	}H_{	24	}}$	&	3800	&&&	$\mathrm {C_{	102	}H_{	26	}}$	&	180	\\
$\mathrm {C_{	27	}H_{	15	}}$	&	2243	&&&	$\mathrm {C_{	51	}H_{	19	}}$	&	3241	&&&	$\mathrm {C_{	58	}H_{	20	}}$	&	3182	&&&	$\mathrm {C_{	87	}H_{	23	}}$	&	649	&&&	$\mathrm {C_{	102	}H_{	26	}}$	&	3214	\\
$\mathrm {C_{	27	}H_{	15	}}$	&	2244	&&&	$\mathrm {C_{	52	}H_{	18	}}$	&	834	&&&	$\mathrm {C_{	59	}H_{	19	}}$	&	89	&&&	$\mathrm {C_{	87	}H_{	25	}}$	&	3271	&&&	$\mathrm {C_{	103	}H_{	27	}}$	&	3285	\\

		\hline
	\end{tabular}
\end{table*}

 \begin{table*}
	\centering
	\caption{PAHdb v3.20 UIDs for cationic PAHs without aliphatic chains.}
	\label{tab:tabA4}
	\begin{tabular}{lcccccccccccccccccc} 
		\hline
		PAH  & UID & & & PAH  & UID & & & PAH  & UID & & & PAH  & UID & & & PAH  & UID \\
		\hline
$\mathrm {C_{	22	}H_{	14	}^{+}}$	&	302	&&&	$\mathrm {C_{	54	}H_{	18	}^{+}}$	&	38	&&&	$\mathrm {C_{	60	}H_{	28	}^{+}}$	&	3714	&&&	$\mathrm {C_{	96	}H_{	25	}^{+}}$	&	664	&&&	$\mathrm {C_{	99	}H_{	25	}^{+}}$	&	3283	\\
$\mathrm {C_{	22	}H_{	14	}^{+}}$	&	306	&&&	$\mathrm {C_{	54	}H_{	17	}^{+}}$	&	39	&&&	$\mathrm {C_{	60	}H_{	28	}^{+}}$	&	3718	&&&	$\mathrm {C_{	96	}H_{	25	}^{+}}$	&	665	&&&	$\mathrm {C_{	102	}H_{	26	}^{+}}$	&	178	\\
$\mathrm {C_{	24	}H_{	12	}^{+}}$	&	19	&&&	$\mathrm {C_{	54	}H_{	20	}^{+}}$	&	3177	&&&	$\mathrm {C_{	60	}H_{	26	}^{+}}$	&	3767	&&&	$\mathrm {C_{	96	}H_{	25	}^{+}}$	&	666	&&&	$\mathrm {C_{	102	}H_{	26	}^{+}}$	&	181	\\
$\mathrm {C_{	24	}H_{	14	}^{+}}$	&	205	&&&	$\mathrm {C_{	54	}H_{	24	}^{+}}$	&	3888	&&&	$\mathrm {C_{	60	}H_{	26	}^{+}}$	&	3769	&&&	$\mathrm {C_{	96	}H_{	25	}^{+}}$	&	667	&&&	$\mathrm {C_{	102	}H_{	26	}^{+}}$	&	3215	\\
$\mathrm {C_{	24	}H_{	14	}^{+}}$	&	207	&&&	$\mathrm {C_{	54	}H_{	24	}^{+}}$	&	3904	&&&	$\mathrm {C_{	60	}H_{	26	}^{+}}$	&	3777	&&&	$\mathrm {C_{	96	}H_{	23	}^{+}}$	&	694	&&&	$\mathrm {C_{	102	}H_{	30	}^{+}}$	&	4275	\\
$\mathrm {C_{	24	}H_{	14	}^{+}}$	&	209	&&&	$\mathrm {C_{	54	}H_{	24	}^{+}}$	&	3905	&&&	$\mathrm {C_{	60	}H_{	26	}^{+}}$	&	3779	&&&	$\mathrm {C_{	96	}H_{	23	}^{+}}$	&	697	&&&	$\mathrm {C_{	102	}H_{	30	}^{+}}$	&	4278	\\
$\mathrm {C_{	27	}H_{	13	}^{+}}$	&	65	&&&	$\mathrm {C_{	54	}H_{	24	}^{+}}$	&	3906	&&&	$\mathrm {C_{	60	}H_{	26	}^{+}}$	&	3781	&&&	$\mathrm {C_{	96	}H_{	22	}^{+}}$	&	700	&&&	$\mathrm {C_{	102	}H_{	30	}^{+}}$	&	4279	\\
$\mathrm {C_{	28	}H_{	14	}^{+}}$	&	729	&&&	$\mathrm {C_{	54	}H_{	22	}^{+}}$	&	3915	&&&	$\mathrm {C_{	60	}H_{	26	}^{+}}$	&	3786	&&&	$\mathrm {C_{	96	}H_{	30	}^{+}}$	&	3341	&&&	$\mathrm {C_{	103	}H_{	27	}^{+}}$	&	3286	\\
$\mathrm {C_{	30	}H_{	14	}^{+}}$	&	153	&&&	$\mathrm {C_{	54	}H_{	22	}^{+}}$	&	3916	&&&	$\mathrm {C_{	60	}H_{	26	}^{+}}$	&	3787	&&&	$\mathrm {C_{	96	}H_{	30	}^{+}}$	&	3354	&&&	$\mathrm {C_{	107	}H_{	27	}^{+}}$	&	3288	\\
$\mathrm {C_{	31	}H_{	15	}^{+}}$	&	3227	&&&	$\mathrm {C_{	54	}H_{	22	}^{+}}$	&	3917	&&&	$\mathrm {C_{	60	}H_{	26	}^{+}}$	&	3801	&&&	$\mathrm {C_{	96	}H_{	30	}^{+}}$	&	3355	&&&	$\mathrm {C_{	107	}H_{	27	}^{+}}$	&	3290	\\
$\mathrm {C_{	32	}H_{	14	}^{+}}$	&	5	&&&	$\mathrm {C_{	54	}H_{	24	}^{+}}$	&	3926	&&&	$\mathrm {C_{	60	}H_{	26	}^{+}}$	&	3802	&&&	$\mathrm {C_{	96	}H_{	30	}^{+}}$	&	3356	&&&	$\mathrm {C_{	108	}H_{	26	}^{+}}$	&	3218	\\
$\mathrm {C_{	32	}H_{	14	}^{+}}$	&	592	&&&	$\mathrm {C_{	55	}H_{	19	}^{+}}$	&	3245	&&&	$\mathrm {C_{	60	}H_{	24	}^{+}}$	&	3812	&&&	$\mathrm {C_{	96	}H_{	30	}^{+}}$	&	3360	&&&	$\mathrm {C_{	110	}H_{	26	}^{+}}$	&	163	\\
$\mathrm {C_{	34	}H_{	16	}^{+}}$	&	126	&&&	$\mathrm {C_{	56	}H_{	20	}^{+}}$	&	3180	&&&	$\mathrm {C_{	60	}H_{	24	}^{+}}$	&	3813	&&&	$\mathrm {C_{	96	}H_{	30	}^{+}}$	&	3361	&&&	$\mathrm {C_{	110	}H_{	30	}^{+}}$	&	184	\\
$\mathrm {C_{	35	}H_{	15	}^{+}}$	&	3230	&&&	$\mathrm {C_{	57	}H_{	19	}^{+}}$	&	644	&&&	$\mathrm {C_{	60	}H_{	24	}^{+}}$	&	3814	&&&	$\mathrm {C_{	96	}H_{	30	}^{+}}$	&	3364	&&&	$\mathrm {C_{	111	}H_{	27	}^{+}}$	&	3292	\\
$\mathrm {C_{	50	}H_{	18	}^{+}}$	&	43	&&&	$\mathrm {C_{	58	}H_{	20	}^{+}}$	&	3183	&&&	$\mathrm {C_{	60	}H_{	24	}^{+}}$	&	3815	&&&	$\mathrm {C_{	96	}H_{	30	}^{+}}$	&	3374	&&&	$\mathrm {C_{	112	}H_{	26	}^{+}}$	&	166	\\
$\mathrm {C_{	51	}H_{	15	}^{+}}$	&	824	&&&	$\mathrm {C_{	59	}H_{	19	}^{+}}$	&	88	&&&	$\mathrm {C_{	60	}H_{	24	}^{+}}$	&	3820	&&&	$\mathrm {C_{	96	}H_{	30	}^{+}}$	&	3375	&&&	$\mathrm {C_{	112	}H_{	26	}^{+}}$	&	607	\\
$\mathrm {C_{	51	}H_{	19	}^{+}}$	&	3242	&&&	$\mathrm {C_{	59	}H_{	21	}^{+}}$	&	3248	&&&	$\mathrm {C_{	96	}H_{	24	}^{+}}$	&	111	&&&	$\mathrm {C_{	96	}H_{	30	}^{+}}$	&	3377	&&&	$\mathrm {C_{	112	}H_{	28	}^{+}}$	&	731	\\
$\mathrm {C_{	52	}H_{	18	}^{+}}$	&	42	&&&	$\mathrm {C_{	60	}H_{	28	}^{+}}$	&	3700	&&&	$\mathrm {C_{	96	}H_{	25	}^{+}}$	&	660	&&&	$\mathrm {C_{	96	}H_{	30	}^{+}}$	&	3378	&&&	$\mathrm {C_{	113	}H_{	27	}^{+}}$	&	3294	\\
$\mathrm {C_{	52	}H_{	16	}^{+}}$	&	825	&&&	$\mathrm {C_{	60	}H_{	28	}^{+}}$	&	3701	&&&	$\mathrm {C_{	96	}H_{	25	}^{+}}$	&	661	&&&	$\mathrm {C_{	98	}H_{	28	}^{+}}$	&	566	&&&	$\mathrm {C_{	115	}H_{	29	}^{+}}$	&	3297	\\
$\mathrm {C_{	52	}H_{	18	}^{+}}$	&	3174	&&&	$\mathrm {C_{	60	}H_{	28	}^{+}}$	&	3711	&&&	$\mathrm {C_{	96	}H_{	25	}^{+}}$	&	662	&&&	$\mathrm {C_{	98	}H_{	28	}^{+}}$	&	568	&&&	$\mathrm {C_{	115	}H_{	27	}^{+}}$	&	3299	\\

		\hline
	\end{tabular}
\end{table*}

\section{Anharmonicity-related tests}\label{sec:app_anharmonicity}

\subsection{Sensitivity to the commonly adopted redshift}\label{subsec:app_redshift}

\begin{figure}
    \centering
    \includegraphics[width=\columnwidth]{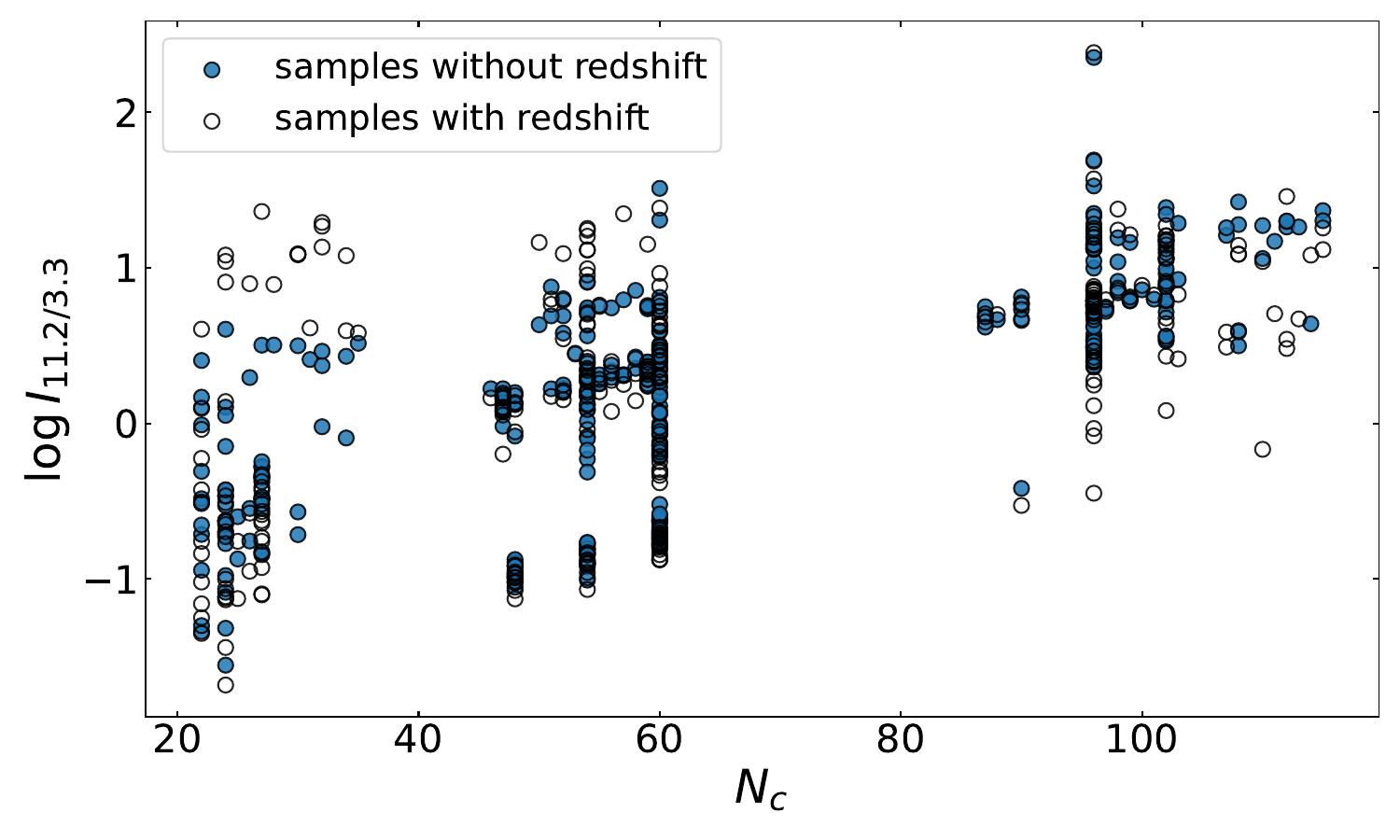}
    \caption{$I_{11.2}/I_{3.3}$ versus  $N_{\mathrm{C}}$, calculated from spectra with and without a $15~\mathrm{cm}^{-1}$ redshift.}
    \label{fig:app_redshift}
\end{figure}

To assess the effect of the commonly adopted anharmonic redshift of $15~\mathrm{cm}^{-1}$, we performed a sensitivity test with and without applying the redshift. We find that the resulting $I_{11.2}/I_{3.3}$ ratios are largely insensitive to this treatment. As shown in Fig.~\ref{fig:app_redshift}, for the vast majority of PAHs, the ratios derived with and without the redshift closely overlap and follow the same overall trend with $N_{\mathrm{C}}$, while any noticeable offsets are confined to the smallest molecules. Therefore, our conclusions are robust against the choice of redshift treatment.

\subsection{Anharmonicity and the $I_{11.2}/I_{3.3}$ diagnostic}\label{subsec:app_ratio}

The 3--4~$\mu$m region, and therefore the  $I_{11.2}/I_{3.3}$ intensity ratio, is known to be particularly sensitive to anharmonicity, temperature effects, and intensity redistribution through overtones and combination bands \cite[e.g.,][]{Mackie2022,2023FaDi..245..380L}.
To assess whether this issue could qualitatively alter the conclusions of the present work, we carried out an external consistency check using the anharmonic PAH spectra computed by \citet{2025MNRAS.541.3073M}, who generated machine-learning molecular dynamics (MLMD) anharmonic IR absorption spectra for PAHs in PAHdb. We matched their database to our sample and identified 104 aromatic species and 47 aliphatic-bearing species with available anharmonic spectra. For the matched subset, we also computed harmonic spectra at 50~K and compared the resulting values of $\log(I_{11.2/3.3})$ between the harmonic and anharmonic cases. Because the data of \citet{2025MNRAS.541.3073M} are absorption spectra, whereas our main analysis is based on emission-cascade spectra, this comparison is intended as a consistency test on the intrinsic sensitivity of the $I_{11.2}/I_{3.3}$ ratio to anharmonicity, rather than as a one-to-one recalibration of the diagnostic grids in the main text.

\begin{figure*}
    \centering
    \includegraphics[width=\textwidth]{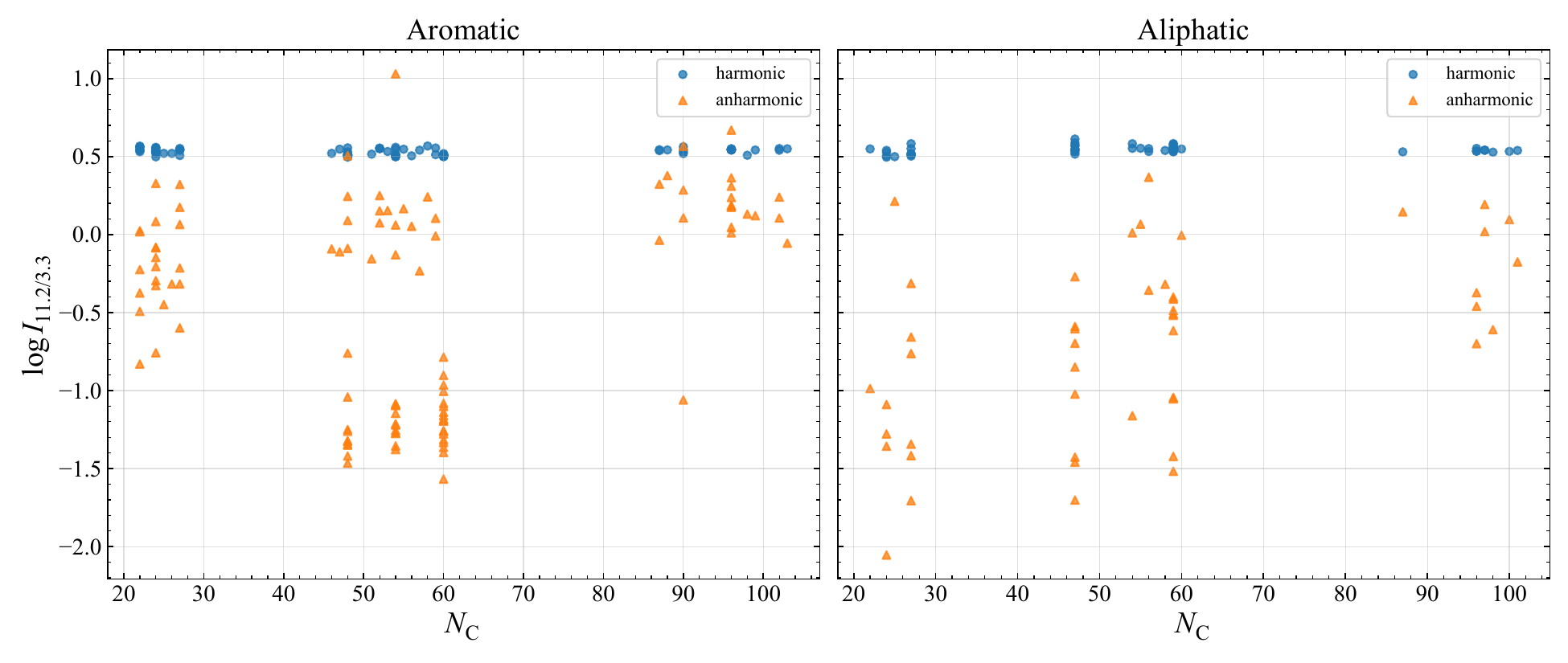}
    \caption{Comparison of harmonic and anharmonic values of $\log(I_{11.2/3.3})$ as a function of carbon number $N_{\mathrm C}$ for the matched subsets of aromatic (left) and aliphatic-bearing (right) PAHs. Blue circles denote the harmonic values computed at 50~K, and orange triangles denote the anharmonic values from \citet{2025MNRAS.541.3073M}. In both subsets, the anharmonic spectra systematically shift the intrinsic $I_{11.2}/I_{3.3}$ ratio relative to the harmonic values.}
    \label{fig:Figure_B2}
\end{figure*}

\begin{figure*}
    \centering
    \includegraphics[width=\textwidth]{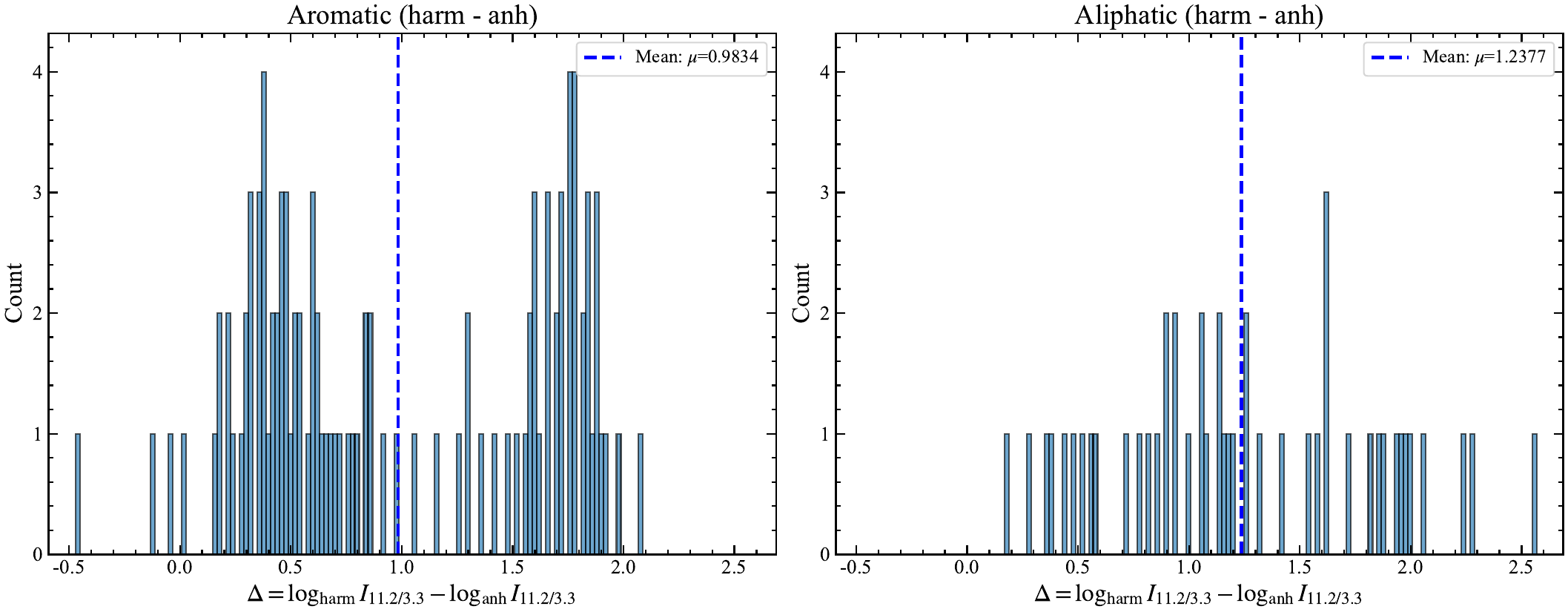}
    \caption{Distribution of $\Delta = \log(I_{11.2/3.3})_{\mathrm{harm}} - \log(I_{11.2/3.3})_{\mathrm{anh}}$ for the matched aromatic (left) and aliphatic-bearing (right) subsets. The dashed blue lines mark the sample means, 0.983 for the aromatic subset and 1.238 for the aliphatic-bearing subset. The positive offsets in both subsets indicate that anharmonicity systematically affects the intrinsic $I_{11.2}/I_{3.3}$ ratio in both aromatic and aliphatic PAHs.}
    \label{fig:Figure_B3}
\end{figure*}

Figure~\ref{fig:Figure_B2} compares the harmonic and anharmonic values of $\log(I_{11.2/3.3})$ as a function of $N_{\mathrm C}$ for the aromatic and aliphatic matched subsets. In both classes, the anharmonic values differ systematically from the harmonic ones, demonstrating that anharmonicity affects the $I_{11.2}/I_{3.3}$ diagnostic for both aromatic and aliphatic PAHs. To quantify this shift, we define
\begin{equation}
\Delta = \log(I_{11.2/3.3})_{\mathrm{harm}} - \log(I_{11.2/3.3})_{\mathrm{anh}}.
\end{equation}
The distributions of $\Delta$ are shown in Figure~\ref{fig:Figure_B3}. For the aromatic subset, we obtain $N=104$, a mean $\Delta$ of 0.983, a median of 0.783, and a 16th--84th percentile range of 0.361--1.768. For the aliphatic-bearing subset, we obtain $N=47$, a mean $\Delta$ of 1.238, a median of 1.160, and a 16th--84th percentile range of 0.560--1.916. 
Thus, the anharmonic correction exhibits the same sign and a comparable order of magnitude in both classes.

Taken together, these results suggest that anharmonicity is more likely to rescale the absolute $I_{11.2}/I_{3.3}$--based size calibration than to erase the relative aromatic--versus--aliphatic trends emphasized in the main text.
In other words, while the absolute carbon-number values inferred from $I_{11.2}/I_{3.3}$ may shift when anharmonic effects are taken into account, the principal conclusion of this paper, that aliphatic components perturb the relative positions of PAH populations within the diagnostic grids, is not qualitatively altered.
The present test also indicates that a more complete anharmonic treatment of aliphatic-bearing PAHs will be important for future attempts to refine the absolute calibration of mixed aromatic--aliphatic PAH diagnostics.

\bsp	
\label{lastpage}
\end{document}